\documentclass[aps,prb,twocolumn,superscriptaddress,floatfix]{revtex4-1}
\usepackage[percent]{overpic}
\usepackage{bm}
\usepackage{graphicx,graphics}
\usepackage{dcolumn}
\newcommand{\vect}[1]{{\bm{#1}}}
\usepackage{amsmath,amssymb,amsfonts}
\usepackage{latexsym,verbatim}
\usepackage{bm}
\usepackage{color}
\usepackage{ulem}
\usepackage[breaklinks=true,colorlinks,citecolor=blue,linkcolor=blue,urlcolor=blue]{hyperref}
\usepackage{braket}

\begin{document}
\title{Quantum non-local theory of topological Fermi arc plasmons in Weyl semimetals}

\author{Gian Marcello Andolina}
\email{gian.andolina@sns.it}
\affiliation{NEST, Scuola Normale Superiore, I-56126 Pisa,~Italy}
\affiliation{Istituto Italiano di Tecnologia, Graphene Labs, Via Morego 30, I-16163 Genova,~Italy}
\author{Francesco M.D. Pellegrino} 
\affiliation{NEST, Scuola Normale Superiore, I-56126 Pisa,~Italy}
\author{Frank H.L. Koppens}
\affiliation{ICREA-Instituci\'{o} Catalana de Recer\c{c}a i Estudis Avancats, Barcelona, Spain}
\affiliation{ICFO-Institut de Ci\`encies Fot\`oniques, The Barcelona Institute of Science and Technology, 08860 Castelldefels (Barcelona), Spain}
\author{Marco Polini}
\affiliation{Istituto Italiano di Tecnologia, Graphene Labs, Via Morego 30, I-16163 Genova,~Italy}

\begin{abstract}
The surface of a Weyl semimetal (WSM) displays Fermi arcs, i.e.~disjoint segments of a two-dimensional Fermi contour. We present a quantum-mechanical non-local theory of chiral Fermi arc plasmons in WSMs with broken time-reversal symmetry. These are collective excitations constructed from topological Fermi arc and bulk electron states and arising from electron-electron interactions, which are treated in the realm of the random phase approximation. Our theory includes quantum effects associated with the penetration of the Fermi arc surface states into the bulk and dissipation, which is intrinsically non-local in nature and arises from decay processes mainly involving bulk electron-hole pair excitations.
\end{abstract}

\maketitle

\section{Introduction} 
Plasmons are self-sustained oscillations of the charge density that occur in metals and doped semiconductors~\cite{Pines_and_Nozieres,Platzman_and_Wolff,Giuliani_and_Vignale}. The coupling of these matter excitations to photons enables to squeeze electromagnetic radiation from the visible~\cite{Maier07} to the Terahertz (THz)~\cite{pablo_naturenano_2017} spectral range into nanoscale devices. However, plasmons suffer scattering from ever-present extrinsic mechanisms in real solid-state devices~\cite{khurgin_naturenano_2015}, including phonons and disorder. Therefore, when confinement is significant, i.e.~when the plasmon wavelength $\lambda_{\rm p}$ in the material is much smaller than the illumination wavelength $\lambda_{0}=2\pi c/\omega$, losses tend to be high (at room temperature) and hamper potential technological breakthroughs. 

Substantial efforts have been recently made to increase the lifetime of these propagating modes at room temperature, without decreasing the associated compression ratio $\lambda_{\rm p}/\lambda_{0}$. For example, one can utilize high-quality graphene sheets encapsulated in hexagonal boron nitride~\cite{geim_nature_2013}, where graphene plasmons scatter essentially only against the acoustic phonons of the two-dimensional (2D) carbon lattice~\cite{woessner_naturemater_2015}, which are weakly coupled to the electronic degrees of freedom. Another possible pathway is to use plasmons in topologically-non-trivial materials~\cite{song_pnas_2016,kumar_prb_2016,dipietro_naturenano_2013,autore_aom_2015}. In the particular case of crystals displaying broken time-reversal symmetry (BTRS), the existence of unidirectional propagating modes akin to the ultra-long-lived~\cite{kumada_prl_2014} topological~\cite{jin_naturecommun_2016} edge magnetoplasmons that occur in 2D 
electron systems in the quantum Hall regime~\cite{Giuliani_and_Vignale} is expected. Technologically, it would be extremely useful to use materials where BTRS occurs {\it without} the aid of an external magnetic field. 
Natural candidates among topological materials with BTRS are recently discovered Weyl semimetals (WSMs)~\cite{Hosur_2013,Gorbar_2016,Hasan_2017,Yan_2017,Burkov_2017,Armitage2018}. These are semimetals with protected linear band crossings in the Brillouin zone, which act as power-law-decaying sources of Berry curvature~\cite{Hosur_2013,Gorbar_2016,Hasan_2017,Yan_2017,Burkov_2017,Armitage2018}. Some of these compounds do display intrinsic BTRS~\cite{Shekhar_arXiv_2016} and, at the same time, have intriguing topological surface states called ``Fermi arcs'' (FAs)~\cite{Hosur_2013,Gorbar_2016,Hasan_2017,Yan_2017,Burkov_2017,Armitage2018}. These are disjoint segments of a 2D Fermi contour---see Fig.~\ref{fig:spectrum}(b)---which have so far been imaged only with angle-resolved photoemission spectroscopy (ARPES)~\cite{xu_science_2015,lv_prx_2015}. Provided that losses are not too strong and that the plasmon electric field leaks sufficiently outside the material, the plasmonic excitations of FA surface states can in 
principle be explored with spatial 
resolution using scanning-type near-field 
optical spectroscopy (for a 
recent 
review see e.g.~Ref.~\onlinecite{basov_science_2016}).

Theoretically, a few aspects of plasmons in WSMs have been studied for both cases of bulk~\cite{pellegrino_prb_2015} and surface~\cite{hofmann_prb_2016} modes.
In particular, the authors of Ref.~\onlinecite{pellegrino_prb_2015}  carried out a study of bulk plasmons in WSMs, by coupling the Maxwell equations as modified by the axion term~\cite{wilczek_prl_1987} with a {\it local} approximation for the constitutive equation, $J_{\alpha}({\bm q}, \omega)\approx \sum_{\beta}\sigma_{\alpha\beta}({\bm 0}, \omega) E_{\beta}({\bm q}, \omega)$. The local conductivity tensor $\sigma_{\alpha\beta}({\bm 0}, \omega)$ was calculated~\cite{pellegrino_prb_2015} by using semiclassical Boltzmann transport theory in the clean limit. The authors of Ref.~\onlinecite{hofmann_prb_2016}  studied WSM surface plasmon polaritons by using the same approach and deep in the retarded limit, i.e.~for surface plasmon wave numbers $|{\bm q}_{\|}| \sim \omega/c$. Finally, the authors of Ref.~\onlinecite{song_arxiv_2017} carried out a semiclassical study of WSM surface plasmons, which is intrinsically valid in the long-wavelength limit. Non-local corrections were heuristically added to the 
semiclassical 
equations of motion. An important open question on the plasmon lifetime remains. In order to quantify this, the quantum-mechanical coupling between Fermi arc surface states and bulk states needs to be evaluated.

In this work, we present a fully quantum-mechanical theory of WSM Fermi arc (FA) plasmons that goes beyond the state-of-the-art~\cite{pellegrino_prb_2015,hofmann_prb_2016,song_arxiv_2017}. The present derivation focuses on the simplest microscopic model Hamiltonian of a (type-I) WSM with BTRS~\cite{Hosur_2013,Gorbar_2016,Hasan_2017,Yan_2017,Burkov_2017} and is based on linear response theory~\cite{Pines_and_Nozieres,Platzman_and_Wolff,Giuliani_and_Vignale} and the random phase approximation (RPA)~\cite{Pines_and_Nozieres,Platzman_and_Wolff,Giuliani_and_Vignale}. We focus on the electrostatic regime, where the plasmon wave number $|{\bm q}_{\|}|$ is much larger than the photon one ($\omega/c$), enabling a great concentration of electromagnetic energy. We claim that quantum {\it non-local} effects are crucial to understand WSM FA plasmon physics. First of all, our theory captures analytically the first non-local correction to the FA plasmon dispersion in the limit $|{\bm q}_{\|}| \ll k_{\rm F}$, where $k_{\rm F}
$ is the 
bulk Fermi wave number. Second, since the FA wavefunctions are in 
contact (i.e.~strong spatial overlap) with a bulk of gapless excitations, FA plasmons are susceptible to Landau damping even at zero temperature and deep in the long-wavelength $|{\bm q}_{\|}| \ll k_{\rm F}$ limit. We quantify this intrinsic dissipation mechanism, which is dominated by processes whereby FA plasmons decay by emitting electron-hole pairs in the bulk. Our calculations on the FA plasmon lifetime pose strict bounds on the observability of certain angular portions of the highly-anisotropic FA plasmon dispersion. 
Indeed, since FA plasmon modes can be strongly overdamped along certain angular directions, it is essential to consider the dispersion relation as well as the angular dependence of the losses. 
Finally, we study the pattern of FA plasmon waves that can be measured by carrying out a scattering-type near-field optical experiment (s-SNOM)~\cite{basov_science_2016} on the surface of a WSM with BTRS.  
In this class of experiments, light is focused on a metallized tip with the aim of launching propagating surface plasmons.
We calculate the screened potential associated with FA plasmons, which clearly shows characteristic features due to their peculiar propagation dynamics. 

Our Article is organized as following. In Sect.~\ref{sect:single_particle} we present the single-particle Hamiltonian and corresponding bulk and surface eigenvalues and eigenstates of a type-I semi-infinite WSM with BTRS. In Sect.~\ref{sect:linear_response_theory} we lay down a general theory of surface plasmons hosted by electron systems occupying a semi-infinite space. In Sect.~\ref{sect:lindhard_WSM} we present a formal expression of the linear density-density response function of a WSM based on the Lehmann representation. We also introduce analytical asymptotic formulae for the same quantity in the relevant limits with respect to the 2D conserved wave vector ${\bm q}_{\|}$ and frequency $\omega$. Sect.~\ref{sect:dispersion} is devoted to the FA plasmon dispersion, while intrinsic damping of these modes is discussed in Sect.~\ref{sect:damping}. Finally, in Sect.~\ref{sect:SNOM} we present a calculation of the spatial pattern of FA plasmon waves that would be seen in a s-SNOM experiment on the surface of a 
WSM with BTRS. A brief summary is reported in Sect.~\ref{sect:summary}. Five appendices report a number of very useful technical details.

\section{Single-particle physics of semi-infinite WSMs}
\label{sect:single_particle}

 In order to describe a three-dimensional (3D) WSM with BTRS and two Weyl nodes separated by a vector $2{\bm b}$ (see Fig.~\ref{fig:spectrum}), we use the following family of Hamiltonians~\cite{Murakami} depending on the parameter $m$:

\begin{figure}[h!]
\centering
\begin{overpic}[width=1\columnwidth]{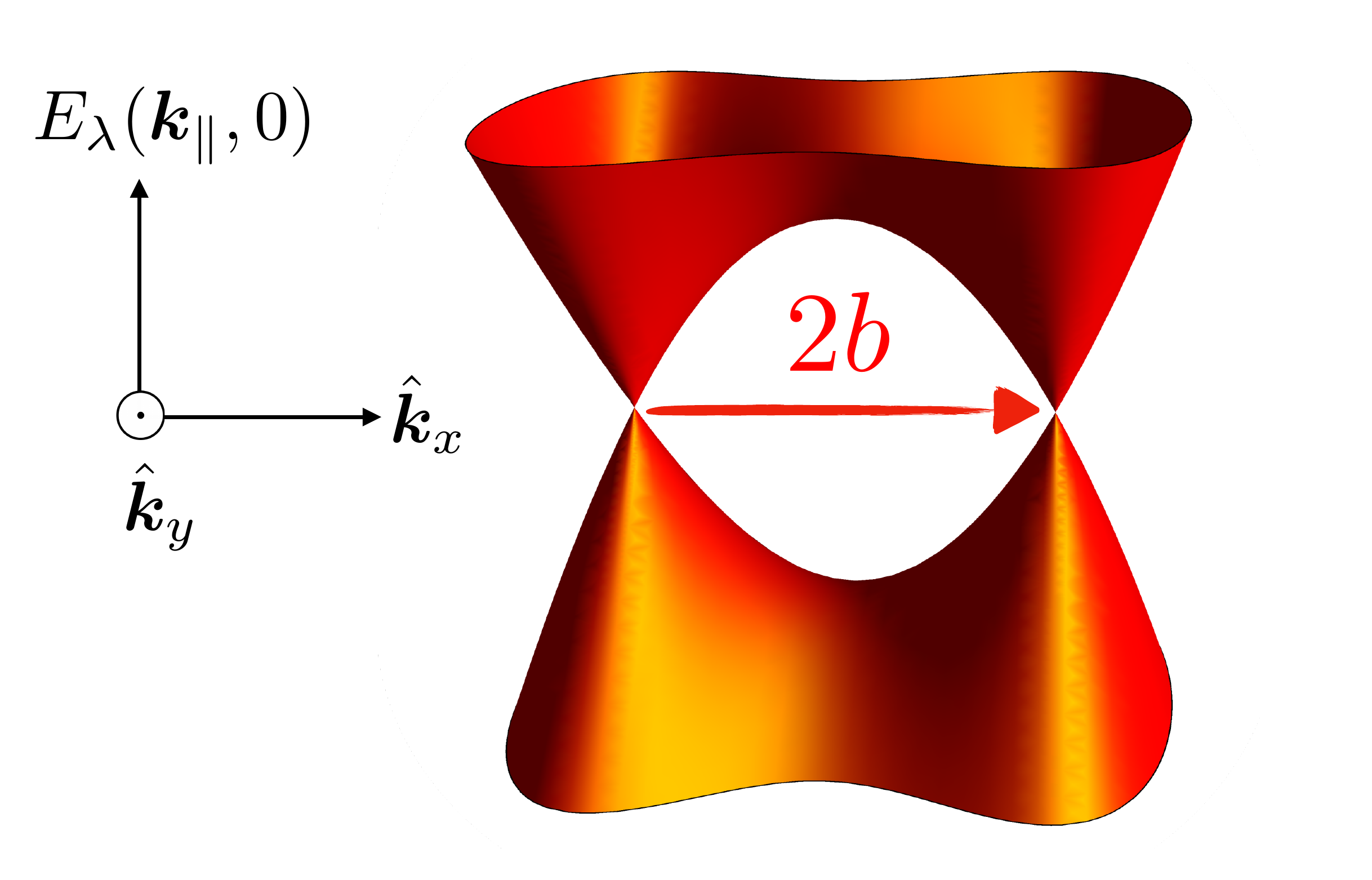}\put(2,70){\normalsize (a)}\end{overpic}\vspace{0.5em}
\begin{overpic}[width=1\columnwidth]{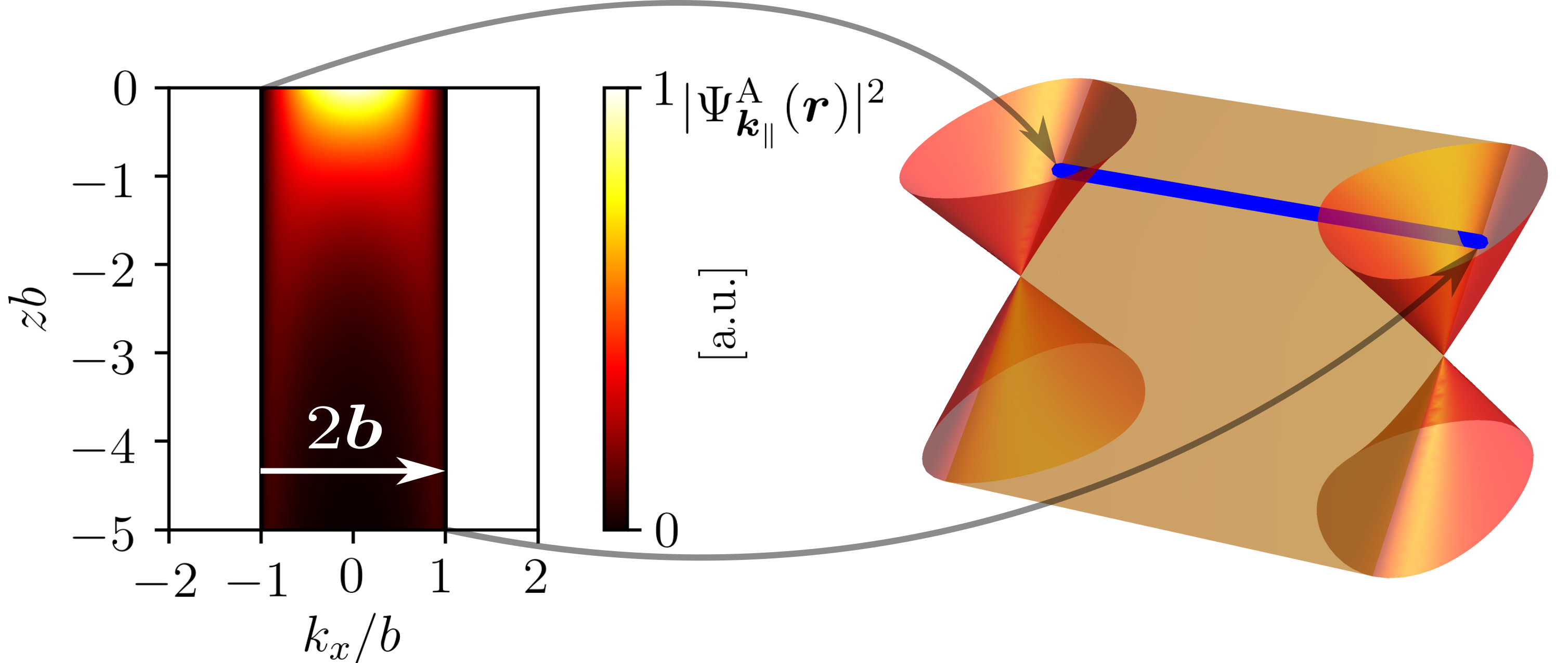}\put(2,42){\normalsize (b)}\end{overpic}\vspace{0.5em}
\caption{(Color online) Panel (a) shows the bulk band structure $E_{\lambda}({\bm k})$ of a WSM with BTRS, for $k_{z}=0$. Note that the two Weyl nodes are separated by a vector $2{\bm b}$ in the $\hat{\bm k}_{x} \| \hat{\bm x}$ direction. Panel (b) shows the square-modulus of the FA wavefunction in Eq.~(\ref{eq:ArcStates}) as a function of the penetration $z$ (in units of $1/b$) into the bulk of a WSM, for a given value of $k_{x}/b$. The right side of the panel shows a low-energy zoom of the bulk band structure with two Weyl cones, with the sheet representing the FA dispersion relation $E_{\rm A}({\bm k}_{\|})$. \label{fig:spectrum}
}
\end{figure}

\begin{equation}\label{eq:hamiltonian}
{\cal H}_{m}(\vect{k})=\frac{\hbar v}{2b}(k_x^2-m)\sigma_x+\hbar v(k_y \sigma_z + k_z \sigma_y)~.
\end{equation}
In Eq.~(\ref{eq:hamiltonian}), $v$ is the Weyl fermion's velocity and $\sigma_i$ are ordinary $2\times2$ Pauli matrices. 
The first term on the right-hand side of Eq.~(\ref{eq:hamiltonian}) breaks time-reversal symmetry. For $m=b^2>0$, the low-energy spectrum of ${\cal H}_{m}(\vect{k})$ contains two Weyl nodes separated by $2{\bm b} = 2b \hat{\bm k}_{x}$ with $b>0$. 
The spectrum of ${\cal H}_{m}(\vect{k})$ is $E_{\lambda}({\bm k})=\lambda \hbar v\sqrt{K^2_{x}+k^2_{y}  + k^2_{z}}$, where $\lambda=\pm1$ denotes conduction/valance band states and $K_{x} \equiv (k_x^2-b^2)/(2b)$. For $k_x \approx \pm b$, $E_{\lambda}({\bm k})$ is linear and isotropic, with two Weyl cones located at $\vect{k}=\pm b \hat{\vect{k}}_{x}$. The corresponding eigenstates are $\Psi_{\lambda,\vect{k}}(\vect{r}) =V^{-1/2} u_{\vect{k},\lambda}e^{i\vect{k}\cdot \vect{r}}$, where $V = S L_{z}$ is the 3D electron system volume, $S$ the surface area, $L_{z}$ its thickness, and $ u_{\vect{k},\lambda}$ is a spinor given by:
\begin{equation}\label{eq:uk}
\begin{split}
u_{\vect{k},\lambda}=\frac{1}{\sqrt{2}}
\begin{pmatrix}
\sqrt{1+\lambda\cos{(\beta_\vect{k})}}e^{-i\phi_\vect{k}}\\
\lambda\sqrt{1-\lambda\cos{(\beta_\vect{k})}}
\end{pmatrix} ~,
\end{split}
\end{equation}
with $\cos{(\beta_\vect{k})}=k_y/\sqrt{K^2_{x} + k^2_{y} + k^2_{z}}$, $\sin{(\beta_\vect{k})}=\sqrt{
(K^2_{x} + k^2_{z})/(K^2_{x} + k^2_{y} + k^2_{z})}$, and $e^{i\phi_\vect{k}} = (K_{x}+ik_z)/\sqrt{K^2_{x}+k^2_{z}}$. 

In this work, we are interested in the surface states of a 3D WSM, which we obtain from the following procedure. We first note that, for $m<0$, ${\cal H}_{m}(\vect{k})$ describes an insulator with a gap $E_{\rm g} = -\hbar v m/(2b)$. The vacuum can be therefore modelled by taking the $\lim_{m\rightarrow-\infty} {\cal H}_{m}(\vect{k})$. Fermi arcs (FAs) are present only on a surface parallel to~\cite{Rao,Yan_2017} $2\vect{b} = 2b\hat{\bm k}_{x}$. Orienting the crystal in such a way that $(\hat{\bm k}_{x}, \hat{\bm k}_{y}, \hat{\bm k}_{z}) \| (\hat{\bm x}, \hat{\bm y}, \hat{\bm z})$, we locate the vacuum/WSM interface hosting the FAs at $z=0$. Surface states emerge by assuming that that the parameter $m$ in Eq.~(\ref{eq:hamiltonian}) changes sign with $z$, i.e.~
$m(z)=b^2\Theta(-z)-\tilde{m} \Theta(z)$, where $\Theta(z)$ is the Heaviside step function and $\tilde{m}$ is a positive constant. This choice describes a WSM for $z<0$ and the vacuum for $z>0$, provided that one takes the limit $\tilde{m}\rightarrow+\infty$. Breaking translational invariance along the $\hat{\bm z}$ direction requires the change $k_z\to -i\partial_z$. FAs are described by states that are bound to the surface. Following Ref.~\onlinecite{Murakami}, we find that their wave functions are 
\begin{equation}
\begin{split}
\label{eq:ArcStates}
\Psi^{\rm A}_{\vect{k}_\parallel}(\vect{r}) =\sqrt{\frac{b^2 - k^2_{x}}{bS}}
		\begin{pmatrix}
			1\\
			0
		\end{pmatrix}\Theta(-z) e^{z/\ell}e^{i\vect{k}_\parallel \vect{r}_\parallel}
 \end{split}
\end{equation}
where $\vect{k}_\parallel = (k_{x}, k_{y})$, $\ell = 2b/({b}^2-k_x^2)$, and the index ``A'' in Eq.~(\ref{eq:ArcStates}) stands for ``arc''. These states have chiral dispersion,  $E_{\rm A}({\bm k}_{\|})=\hbar vk_y$, group velocity given by $v$, density of states given by ${\cal N}_{\rm A} = b/(2\pi^2\hbar v)$ (which is independent of the bulk doping), and exist only between the two Weyl nodes, i.e.~for $-b\leq k_{x}\leq b$. Note that $\ell$, which physically represents the extension of the FA state into the WSM bulk $z<0$, goes to infinity for $k_{x}\to \pm b$. This means that for $k_{x}$ near the location of the Weyl nodes, leakage of the FA states into the bulk cannot be neglected. For the sake of simplicity, the FA states (\ref{eq:ArcStates}) will be denoted by the shorthand $\ket{\rm A}$. 

The presence of the WSM/vacuum interface at $z=0$ affects also the propagating (i.e.~bulk) states. Since 
$\bm{k}_{\|}$ is a good quantum number (because of translational invariance in the $\hat{\bm x}$-$\hat{\bm y}$ plane), we obtain propagating states in the presence of the interface by taking linear combinations of bulk plane-wave states with positive ($k_{z}>0$) and negative ($-k_{z}<0$) values of the $\hat{\bm z}$ component of the wave vector ${\bm k}$. Assuming specular reflection at the interface, we find
\begin{equation}\label{eq:eigenvectorsReflected}
\Psi^{\rm B}_{\lambda,\vect{k}}(\vect{r}_{\|},z) =
\frac{\Theta(-z)}{\sqrt{V}}( u_{\vect{k},\lambda} e^{ik_zz}+r_\vect{k} u_{\vect{\bar{k}},\lambda} e^{-ik_zz})e^{i\vect{k}_\parallel \cdot \vect{r}_\parallel}~,
\end{equation}
where $\vect{\bar{k}}=(\vect{k}_\parallel,-k_z)$ is the reflected wave vector  and $r_{\bm k}$ is a reflection coefficient (with $|r_{\bm k}|=1$) to be determined by imposing suitable boundary conditions. Of course, $\Psi^{\rm B}_{\lambda,\vect{k}}(\vect{r}_{\|},z)$ is an eigenstate of the half-space Hamiltonian $\lim_{\tilde{m}\to +\infty}{\cal H}_{m(z)}({\bm k}_{\|}, k_{z}\to -i\partial_z)$ with eigenvalue $E_{\lambda}({\bm k})$. As explained in Appendix~\ref{app:BCs} , the continuity of the wave functions at the interface requires $r_{\bm k}=-1$. Because the ``reflected'' states (\ref{eq:eigenvectorsReflected}) originate from the WSM bulk states $\Psi_{\lambda,\vect{k}}(\vect{r})$, we will denote them by the shorthand $\ket{\rm B}$. 

\section{Surface plasmons of semi-infinite WSMs}
\label{sect:linear_response_theory}

Due to the presence of the interface at $z=0$, it is natural to lay down linear response theory~\cite{Giuliani_and_Vignale} by imposing translational invariance in the direction perpendicular to $\hat{\bm z}$ and that the wave vector $\vect{q}_\parallel$ parallel to the surface is a good quantum number. The linear density response $n_1$ induced by an external potential is therefore a function of $z$, $\vect{q}_\parallel$, and frequency $\omega$. It can be expressed in terms of the screened potential~\cite{Giuliani_and_Vignale}, $V_{\rm sc}(z,\vect{q}_\parallel,\omega)$, and the proper density-density response function~\cite{Giuliani_and_Vignale} $\tilde{\chi}_{nn}(z, z^\prime, \vect{q}_\parallel,\omega)$ according to 
\begin{equation}\label{eq:Density}
n_1(z,\vect{q}_\parallel,\omega) = \int_{-\infty}^{0} dz^\prime\tilde{\chi}_\textit{nn}(z,z^\prime,\vect{q}_\parallel,\omega) V_{\rm sc}(z^\prime,\vect{q}_\parallel,\omega)~.
\end{equation}
For a self-sustained oscillation occurring in the absence of an external potential, 
the screened potential is related to the induced density by
\begin{eqnarray}\label{eq:Potential}
V_{\rm sc}(z,\vect{q}_\parallel,\omega) &=& \int dz^\prime \int dz^{\prime\prime}v(z,z^\prime,\vect{q}_\parallel)\tilde{\chi}_{nn}(z^\prime,z^{\prime\prime},\vect{q}_\parallel,\omega) \nonumber \\
&\times &V_{\rm sc}(z^{\prime\prime},\vect{q}_\parallel,\omega)~,
\end{eqnarray}
where the integrals over $z^{\prime}$ and $z^{\prime\prime}$ span the half-space $z<0$ and $v(z,z^\prime,\vect{q}_\parallel)= 2\pi e^2 \exp{(-q_\parallel | z-z^\prime|)}/q_{\parallel}$ is the Fourier transform of the Coulomb potential ($q_\parallel=|\vect{q}_\parallel |$). 

Plasmons are non-trivial solutions of the integral equation (\ref{eq:Potential}). In this work we are solely interested in plasmons whose associated electric field is bound to the WSM/vacuum interface. We therefore set $V_{\rm sc}(z,\vect{q}_\parallel,\omega)={\bar v}_{\rm sc}(\vect{q}_\parallel,\omega)e^{-q_\parallel |z|}$, which describes a mode bound to the surface with a ``localization'' length scale equal to $1/q_\parallel$. Utilizing this Ansatz for $V_{\rm sc}(z,\vect{q}_\parallel,\omega)$, we evaluate~\cite{pitarke_rpp_2007} Eq.~(\ref{eq:Potential}) for $z>0$. We find that this integral equation can be written as a standard algebraic 2D plasmon equation,
\begin{equation}\label{eq:plasmon_2D}
1=\frac{2\pi e^2}{q_\parallel}\chi_{\rm eff}(\vect{q}_\parallel,\omega)~,
\end{equation}
provided that one introduces the effective 2D proper response function:
\begin{eqnarray}\label{eq:effective_2D}
\chi_{\rm eff}({\bm q}_\parallel,\omega)&\equiv&\int_{-\infty}^0dz\int_{-\infty}^0dz^\prime \tilde{\chi}_{nn}(z,z^{\prime},\vect{q}_\parallel,\omega) e^{q_\parallel(z+z^\prime)} \nonumber\\
&=& L_{z} \tilde{\chi}_{nn}(q_z,q_z^{\prime},\vect{q}_\parallel,\omega)|_{q_z=iq_\parallel,q_z^{\prime}=-iq_\parallel}~.
\end{eqnarray}
Here,
\begin{eqnarray}\label{eq:3D_proper_response_momentum_space}
\tilde{\chi}_{nn}(q_z,q_z^{\prime},\vect{q}_\parallel,\omega)&\equiv& \frac{1}{L_{z}}\int \frac{dq_z}{2\pi}\int \frac{dq^\prime_z}{2\pi}\tilde{\chi}_{nn}(z,z^{\prime},\vect{q}_\parallel,\omega) \nonumber\\
&\times &e^{-iq_zz} e^{+iq^\prime_zz^\prime}~.
\end{eqnarray}
No approximation has yet been made on $\tilde{\chi}_{ nn}(q_z,q_z^{\prime},\vect{q}_\parallel,\omega)$. The only key assumption we used to derive Eq.~(\ref{eq:plasmon_2D}) was that $\tilde{\chi}_{nn}(z,z^{\prime},\vect{q}_\parallel,\omega)\propto \Theta(-z)\Theta(-z^\prime)$.  

Screening due to the dielectric background can be taken care of by replacing $e^2 \to e^2/\bar{\epsilon}$ in Eq.~(\ref{eq:plasmon_2D}), where $\bar{\epsilon}\equiv (1+\epsilon_{\rm b})/2$ and $\epsilon_{\rm b}$ is the high-frequency WSM bulk dielectric constant. This prescription, which is valid only for surface modes, is demonstrated in Appendix~\ref{app:screening}.

\section{Non-interacting response function of a semi-infinite WSM}
\label{sect:lindhard_WSM}

In the RPA~\cite{Giuliani_and_Vignale}, the exact proper response function  $\tilde{\chi}_{nn}(q_z,q_z^{\prime},\vect{q}_\parallel,\omega)$ in Eq.~(\ref{eq:3D_proper_response_momentum_space}) is replaced by the non-interacting response function ${\chi}^{(0)}_{nn}(q_z,q_z^{\prime},\vect{q}_\parallel,\omega)$, which can be evaluated with the aid of the Lehmann representation~\cite{Giuliani_and_Vignale}:
\begin{eqnarray}\label{eq:Lehmann}
\chi^{(0)}_{nn}(q_z,q_z^{\prime},\vect{q}_\parallel,\omega) &=&
\frac{1}{V\hbar} \sum_{m,n} \frac{f_n - f_m}{\omega_{n,m}+\omega+i0^{+}}\braket{n| \hat{n}_{\vect{q}_\parallel,q_z} |m}\nonumber\\
&\times&
\braket{m|\hat{n}_{-\vect{q}_\parallel,-q^\prime_z}|n}~.
\end{eqnarray}
Here, $\ket{n}$, $\ket{m}$ ($E_{n}$, $E_{m}$) denote the eigenstates (eigenenergies) of the non-interacting Hamiltonian, 
$\hbar\omega_{n,m} = E_{n} - E_{m}$ are the excitation energies, $f_{n}$ are the occupation numbers, and $\hat{n}_{\vect{q}_\parallel,q_z}\equiv e^{-i\vect{q}_{\|} \cdot \hat{\vect{r}}_\parallel-
  iq_z\hat{\bm z}}$. As we have seen above, the states of our semi-infinite WSM comprise the FA states (\ref{eq:ArcStates}) and the reflected states (\ref{eq:eigenvectorsReflected}). We can therefore naturally decompose $\chi^{(0)}_{nn}(q_z,q_z^{\prime},\vect{q}_\parallel,\omega)$ into the sum of three contributions:
\begin{equation}
{\chi}^{(0)}_{nn} = {\chi}^{(0)}_{\rm AA}+{\chi}^{(0)}_{\rm BB}+{\chi}^{(0)}_{\rm AB}~.
 \end{equation}
In the first contribution, ${\chi}^{(0)}_{\rm AA}$, the sum over $n$ and $m$ in Eq.~(\ref{eq:Lehmann}) spans only the FA states (\ref{eq:ArcStates}), i.e.~$\ket{n}=\ket{A}$, $\ket{m}=\ket{A}^\prime$. In the second contribution, ${\chi}^{(0)}_{\rm BB}$, the sum spans only the reflected states (\ref{eq:eigenvectorsReflected}), i.e.~$\ket{n}=\ket{B}$, $\ket{m}=\ket{B}^\prime$. 
Finally, the third contribution, ${\chi}^{(0)}_{\rm AB}$, takes into account the remaining cross processes in which $\ket{n}=\ket{A}$ and $\ket{m}=\ket{B}$ or $\ket{n}=\ket{B}$ and $\ket{m}=\ket{A}$. Because of Eq.~(\ref{eq:effective_2D}), the same decomposition holds true for $\chi_{\rm eff}({\bm q}_\parallel,\omega)$, i.e.~$\chi_{\rm eff}=\chi^{\rm eff}_{\rm AA}+\chi^{\rm eff}_{\rm BB}+\chi^{\rm eff}_{\rm AB}$. 
The FA plasmon dispersion is controlled by ${\rm Re}[\chi_{\rm eff}(\vect{q}_\parallel,\omega)]$, while the plasmon damping rate depends on dissipation, i.e.~${\rm Im}[\chi_{\rm eff}(\vect{q}_\parallel,\omega)]$.

To make analytical progress, we concentrate our attention on FA plasmons in the long-wavelength limit, defined by the regime in which $q_\parallel$ is the smallest wave vector scale in the problem. Eq.~(\ref{eq:effective_2D}) implies $q_z\sim q_z^\prime\sim q_\parallel$. We denote by the shorthand $q$ the small quantity $q_z\sim q_z^\prime\sim q_\parallel$. 

In the limit $q \ll b, \omega/v$, we find that the leading order FA effective response function reduces to (Appendix~\ref{app:AA_response})
\begin{eqnarray}\label{eq:AA1}
{\rm Re}[\chi^{\rm eff}_{\rm AA}]&\to& \frac{q_\parallel \cos(\theta)}{\hbar\omega(2\pi)^2}\Bigg[ 2b\left(1+\frac{vq_\parallel \cos(\theta)}{\omega} \right) \nonumber  \\
&-&q_\parallel |\sin(\theta)| \Bigg]~.
\end{eqnarray}
Like the FA density-of-states ${\cal N}_{\rm A}$, ${\rm Re}[\chi^{\rm eff}_{\rm AA}]$ does not depend on the bulk carrier density $n$.  Neglecting for a moment the contribution to Eq.~(\ref{eq:plasmon_2D}) from the reflected states, ${\rm Re}[\chi^{\rm eff}_{\rm BB}]$, and the cross term, ${\rm Re}[\chi^{\rm eff}_{\rm AB}]$, we find a long-wavelength FA plasmon dispersion
\begin{equation}\label{eq:FA_partial}
\Omega^{\rm FA}_{\theta}=\frac{\alpha_{\rm ee}vb }{ \pi}\cos(\theta)~,
\end{equation}
where $\theta$ is the angle between ${\bm q}_{\|}$ and the $\hat{\bm y}$ axis (i.e.~the direction along which the FA states have finite group velocity), $\cos(\theta) =q_{y}/ q_{\|}$, and $\alpha_{\rm ee} = e^2/(\hbar v\bar{\epsilon})$ is a dimensionless coupling constant describing the strength of electron-electron interactions. The strong directionality dependence of $\Omega^{\rm FA}_{\theta}$ is evident. Interestingly, the group velocity ${\bm v}^{\rm FA}_{\theta} = \hat{\theta} q^{-1}_{\|}\partial_{\theta} \Omega^{\rm FA}_{\theta} =  -\alpha_{\rm ee} v b \sin(\theta) \hat{\theta}/(\pi q_{\|})$ of the mode in Eq.~(\ref{eq:FA_partial}) vanishes when the plasmon wave vector ${\bm q}_{\|}$ is in the ${\hat{\bm y}}$ direction.

We now turn to evaluate the contribution ${\chi}^{(0)}_{\rm BB}(q_z,q_z^{\prime},\vect{q}_\parallel,\omega)$ due to the reflected states (\ref{eq:eigenvectorsReflected}). For a Fermi wave number $k_{\rm F} \ll b$, we can linearize the spectrum $E_{\pm}({\bm k})$ around the points $k^{(j)}_{x}=j b$, where $j=\pm1$ is the Weyl cone index. Again, we work in the long-wavelength limit, which in this case is defined by $q_\parallel \ll \omega/v, k_{\rm F}$. Noting that $q_z, q_z^\prime \sim q_\parallel$, we can state that $q_\parallel$ is the smallest wave vector in the problem. Moreover, we assume to be deep in the single-particle optical absorption gap, i.e.~we consider $\hbar \omega \ll 2E_{\rm F}$ where $E_{\rm F} = \hbar v k_{\rm F}$ is the Fermi energy. In this regime, inter-band contributions to all response functions involving the reflected states (\ref{eq:eigenvectorsReflected}) can be neglected and we can safely use the high-frequency moment expansion~\cite{Giuliani_and_Vignale,torre_prb_2017} for the 
intra-band 
contribution to ${\rm Re}[{\chi}^{(
0)}_{\rm BB}]$. As detailed in Appendix~\ref{app:BB_response}, we find
\begin{equation}\label{eq:GBB}
{\rm Re}[\chi^{\rm eff}_{\rm BB}(\vect{q}_\parallel,\omega)] \to \frac{nq_\parallel}{m_{\rm eff}\omega^2}~,
\end{equation}
where $n = k^3_{\rm F}/(6\pi^2)$ is the density of bulk electrons and $m_{\rm eff}= E_{\rm F}/v^2$ the effective mass. Corrections to Eq.~(\ref{eq:GBB}) scale as $q^3_{\|}$, while terms $\propto q^2_{\|}$ are exactly zero. Notice that inserting only ${\rm Re}[\chi^{\rm eff}_{\rm BB}(\vect{q}_\parallel,\omega)]$ in Eq.~(\ref{eq:plasmon_2D}), we recover the well-known Ritchie relation~\cite{ritchie_pr_1957} $\Omega_{\rm s} = \Omega_{\rm b}\sqrt{\epsilon_{\rm b}/(2\bar{\epsilon})} = v k_{\rm F}\sqrt{\alpha_{\rm ee}/(3\pi)}$  for the surface plasmon frequency,  $\Omega^2_{\rm b} = 4\pi n e^2/(\epsilon_{\rm b} m_{\rm eff})$ being the square of the bulk plasmon frequency. In the undoped $k_{\rm F}=0$ limit, Eq.~(\ref{eq:GBB}) yields ${\rm Re}[\chi^{\rm eff}_{\rm BB}(\vect{q}_\parallel,\omega)]=0$.

Finally, we need to evaluate the cross contributions ${\chi}^{(0)}_{\rm AB}(q_z,q_z^{\prime},\vect{q}_\parallel,\omega)$ and ${\rm Re}[\chi^{\rm eff}_{\rm AB}(\vect{q}_\parallel,\omega)]$. Up to leading order in the long-wavelength limit and, as above, deep in the single-particle optical absorption gap $\hbar\omega/(2E_{\rm F})\ll 1$, we find (Appendix~\ref{app:AB_response})
\begin{equation}\label{eq:GAB}
{\rm Re}[\chi^{\rm eff}_{\rm AB}(\vect{q}_\parallel,\omega)] \to -\frac{q^2_\parallel \big[1+\sin^2(\theta)\big]}{3\pi^2}\frac{1}{\sqrt{2E_{\rm F} |\hbar \omega|}}~.
\end{equation}
Clearly Eq.~(\ref{eq:GAB}) cannot be used in the undoped $k_{\rm F} \to 0$ limit. In this case an explicit calculation ( Appendix~\ref{app:AB_response}) yields $\lim_{k_{\rm F} \to 0} {\rm Re}[\chi^{\rm eff}_{\rm AB}(\vect{q}_\parallel,\omega)]=0$. Note that $\chi^{\rm eff}_{\rm AB}(\vect{q}_\parallel,\omega)$ does not depend on $b$ to leading order for $q_{\|}\ll b$.  In this limit, indeed, $\chi^{\rm eff}_{\rm AB}(\vect{q}_\parallel,\omega)$ describes electron-hole transitions near each Weyl node.

\section{FA plasmon dispersion}
\label{sect:dispersion}

\begin{figure}[t]
\centering
\begin{overpic}[width=\columnwidth]{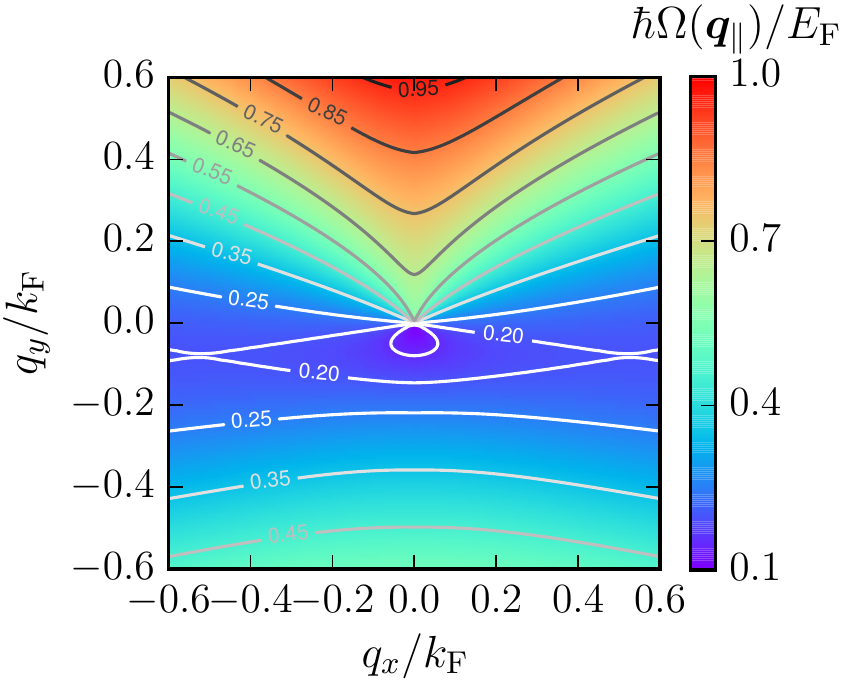}\put(2,72){\normalsize (a)}\end{overpic}\vspace{0.5em}
\begin{overpic}[width=\columnwidth]{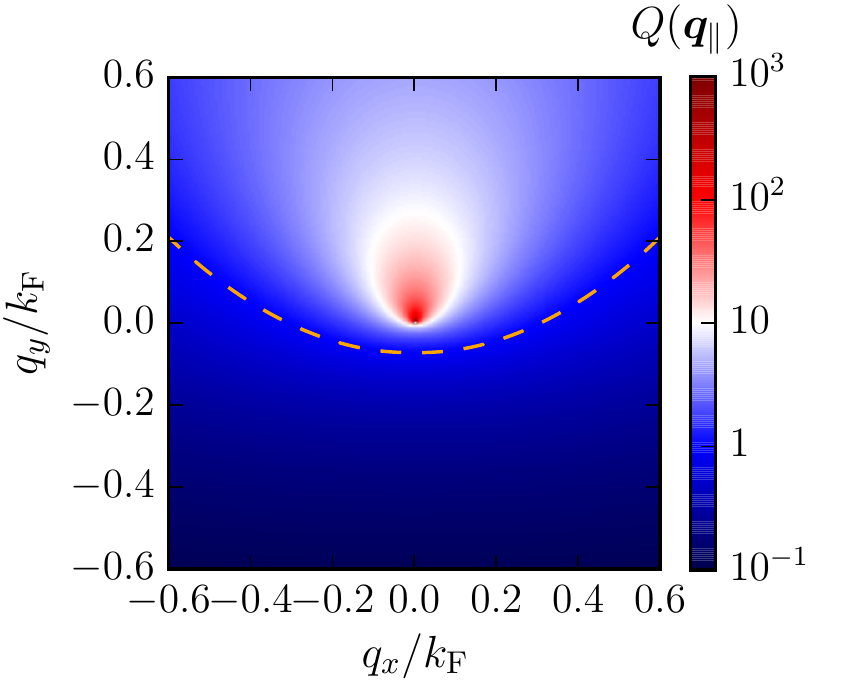}\put(2,72){\normalsize (b)}\end{overpic}
\caption{(Color online) Panel (a) displays the FA plasmon dispersion relation $\Omega(\vect{q}_\parallel)$ in units of $E_{\rm F}/\hbar$ and as a function of $q_{x}/k_{\rm F}$ and $q_{y}/k_{\rm F}$. The plasmon group velocity $\bm{v}_{\rm g}$ is orthogonal to the contour lines. Panel (b) shows the quality factor $Q(\vect{q}_\parallel)$ as a function of $q_{x}/k_{\rm F}$ and $q_{y}/k_{\rm F}$. The orange dashed line represents $Q(\vect{q}_\parallel)=1$. Below this line the plasmon is not a well-defined excitation. In both panels we have set $\alpha_{\rm ee}=0.5$ and $b=3 k_{\rm F}$.\label{fig:two}}
\end{figure}
We now solve Eq.~(\ref{eq:plasmon_2D}) in the RPA, i.e.~by replacing ${\rm Re}[\chi_{\rm eff}(\vect{q}_\parallel,\omega)]$ with ${\rm Re}[\chi^{\rm eff}_{\rm AA}(\vect{q}_\parallel,\omega) +\chi^{\rm eff}_{\rm BB}(\vect{q}_\parallel,\omega) + \chi^{\rm eff}_{\rm AB}(\vect{q}_\parallel,\omega)]$. In the following, we focus only on positive frequencies, since modes with negative frequency can be obtained from the relation $\Omega(\vect{q}_\parallel )=-\Omega(-\vect{q}_\parallel)$. 

\begin{figure}[t]
\centering
\begin{overpic}[width=\columnwidth]{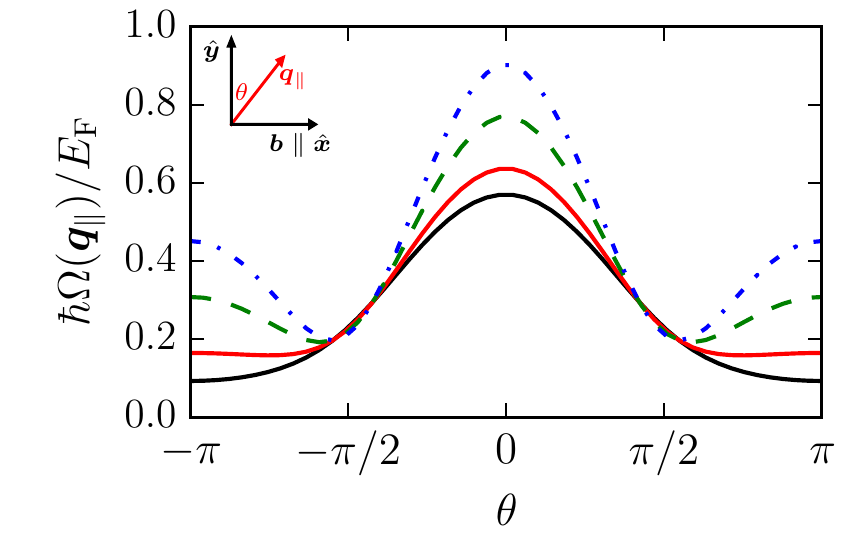}\put(2,52){\normalsize (a)}\end{overpic}\vspace{0.5em}
\begin{overpic}[width=\columnwidth]{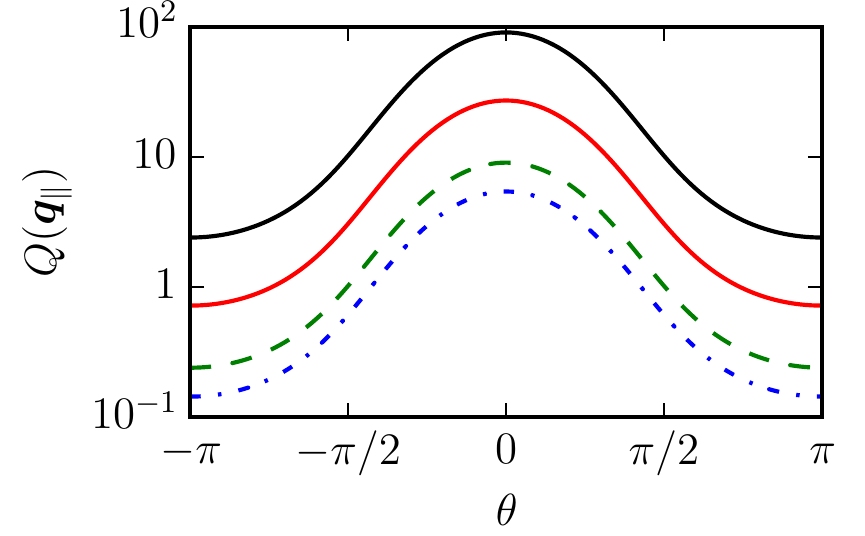}\put(2,52){\normalsize (b)}\end{overpic}
\caption{(Color online) 
Panel (a) The plasmon frequency $\Omega(\vect{q}_{\parallel})$ in units of $E_{\rm F}/\hbar$ as a function of $\theta$ for different values of  $q_{\parallel}= |{\bm q}_{\|}|$ (black solid line: $q_{\parallel} = 0$; red solid line: $q_{\parallel} =0.1~k_{\rm F}$; green dashed line: $q_{\parallel} =0.3~k_{\rm F}$; 
blue dash-dotted line: $q_\parallel =0.5~k_{\rm F}$). The inset shows the definition of $\theta$, i.e.~the angle between the group velocity of the FA states and the wave vector ${\bm q}_{\|}$.
Panel (b) The dimensionless plasmon quality factor $Q(\vect{q}_{\parallel})$  as a function of $\theta$ for different values of  $q_{\parallel}$ (black solid line: $q_{\parallel} =0.03~k_{\rm F}$; red solid line: $q_{\parallel} =0.1~k_{\rm F}$, green dashed line: $q_\parallel =0.3~k_{\rm F}$, blue dash-dotted line: $q_\parallel = 0.5~k_{\rm F}$). In both panels we have set $\alpha_{\rm ee}=0.5$ and $b =3~k_{\rm F}$.\label{fig:three}}
\end{figure}

For doped WSMs ($k_{\rm F} \neq 0$) we find that the plasmon dispersion in the long-wavelength limit is given by $\Omega(\vect{q}_\parallel)=\Omega_{\theta} + \delta \Omega(\vect{q}_\parallel)$, where $\Omega_{\theta}$ is the value of the plasmon frequency for $q_\parallel \rightarrow 0$ and $\delta \Omega(\vect{q}_\parallel)$ is the first-order non-local correction in $q_\parallel$:
\begin{equation}\label{eq:OmegaP}
\Omega_\theta= \frac{1}{2} \Big[\Omega_\theta^{\rm FA}  + \sqrt{(\Omega^{\rm FA}_{\theta})^2 +4\Omega^2_{\rm s}}\Big]
\end{equation}
and 
\begin{equation}
\label{eq:OmegaP1}
\delta \Omega(\vect{q}_\parallel)=\alpha_{\rm ee}vq_\parallel \mathcal{I}(\theta)~.
\end{equation}
In Eq.~(\ref{eq:OmegaP1}) we have introduced the quantity
\begin{eqnarray}\label{eq:I}
\mathcal{I}(\theta) &=&\frac{\Omega_{\theta}}{ \sqrt{(\Omega^{\rm FA}_{\theta})^2 +4\Omega^2_{\rm s}}} \Bigg\{\frac{\cos(\theta)}{2\pi}\left[\frac{2bv\cos(\theta)}{\Omega_\theta}-|\sin(\theta)|\right] \nonumber\\
&-&\frac{2[1+\sin^2(\theta)]}{3\pi}\sqrt{\frac{\hbar \Omega_\theta}{2E_{\rm F}}}~ \Bigg\}.
\end{eqnarray}
The first line in the previous equation is due to ${\rm Re}[\chi^{\rm eff}_{\rm AA}(\vect{q}_\parallel,\omega)]$, while the second line is due to ${\rm Re}[\chi^{\rm eff}_{\rm AB}(\vect{q}_\parallel,\omega)]$. Eq.~(\ref{eq:OmegaP}) is formally identical~\cite{hofmann_prb_2016} to the dispersion relation of a surface magnetoplasmon~\cite{wallis_prb_1974} in the case of a magnetic field oriented along ${\bm b}$ and with $\Omega^{\rm FA}_{\theta} \to \omega_{\rm c} \cos(\theta)$, $\omega_{\rm c}$ being the cyclotron frequency. This result reflects the fact that the topological FA states play the same role of edge states in the quantum Hall regime. The FA plasmon dispersion relation $\Omega({\bm q}_{\|})$ is illustrated in Figs.~\ref{fig:two}(a) and~\ref{fig:three}(a). In Fig.~\ref{fig:two}(a) we clearly note that, for every value of $q_{x}$, there is a large asymmetry in the dispersion $\Omega({\bm q}_{\|})$ as a function of $q_{y}$. This stems from the fact that, for $q_{y}>0$, $\Omega({\bm q}_{\|})$ is 
dominated by the FA states. On the other hand, for $q_{y}<0$, the main contribution to $\Omega({\bm q}_{\|})$ is given by $\Omega_{\rm s}$, which stems from the reflected states (\ref{eq:eigenvectorsReflected}) and is weakly dispersive. The weak dispersive features seen for $q_{y}<0$ reflect an hybridization between the FA and bulk channels. This mixing of channels with very different symmetries (the former highly directional, while the latter is isotropic) leads to a pair of saddle points in the dispersion relation, which are seen from the contours in Fig.~\ref{fig:two}(a) at $ \hbar \Omega({\bm q}_{\|})/E_{\rm F} \approx 0.20$. In Fig.~\ref{fig:three}(a) we show cuts of $\Omega({\bm q}_{\|})$, each one taken at a fixed value of $q_{\|}$, and plotted as a function of $-\pi \leq \theta \leq \pi$.

For typical experimental values~\cite{Armitage2018}, 
$\epsilon_{\rm b}=10$, $E_{\rm F}=40~{\rm meV}$, $b=0.05~{\rm  {\AA}}^{-1}$, and $v = c/1000$, Eq.~(\ref{eq:OmegaP}) yields a FA plasmon energy $\Omega_{\theta}$ in the forward $\theta \approx 0$ direction on the order of $30~{\rm meV}$, corresponding to a frequency $\nu_{\theta} \equiv \Omega_{\theta}/(2\pi)$ of $\sim 7.3~{\rm THz}$. 

In the undoped $k_{\rm F}\to 0$ limit the plasmon dispersion is
\begin{equation}\label{eq:Omegazerodoping}
\Omega(\vect{q}_\parallel)=\Omega_\theta^{\rm FA} +q_{\|} \left[v\cos(\theta)-|\sin(\theta)|\frac{\Omega_\theta^{\rm FA}}{2b}\right]~.
\end{equation}
We will momentarily show that in the undoped limit the FA plasmon is strongly damped, but for non-zero $k_{\rm F}$, this damping is suppressed.

\section{Intrinsic lifetime of FA plasmons}
\label{sect:damping}

So far, we have treated the plasmon as a solution of Eq.~(\ref{eq:Potential}) occurring on the real frequency axis. For fundamental reasons related to causality~\cite{Giuliani_and_Vignale}, however, retarded response functions must have poles located below the real axis, i.e.~at $\omega=\Omega(\vect{q}_\parallel) - i \Gamma(\vect{q}_\parallel)$ with $\Gamma(\vect{q}_\parallel)>0$. When the imaginary part is small, i.e.~when $\Gamma(\vect{q}_\parallel)\ll \Omega(\vect{q}_\parallel)$, the plasmon is a well-defined collective excitation of the many-body system~\cite{Giuliani_and_Vignale}. The plasmon lifetime $\tau({\bm q}_{\|})$ is $\Gamma^{-1}({\bm q}_{\|})$. Expanding Eq.~(\ref{eq:plasmon_2D}) in the small parameter $\Gamma(\vect{q}_\parallel)/ \Omega(\vect{q}_\parallel)$ we obtain:
\begin{equation}\label{eq:LifeTime}
\Gamma(\vect{q}_\parallel)= \frac{{\rm Im}[\chi_{\rm eff}(\vect{q}_\parallel,\Omega(\vect{q}_\parallel)]}{\partial_\omega {\rm Re}[\chi_{\rm eff}(\vect{q}_\parallel,\omega)]|_{\omega=\Omega(\vect{q}_\parallel)}}~.
\end{equation}
We define the quality factor as $Q(\vect{q})\equiv \Omega(\vect{q}_\parallel)/[2\Gamma(\vect{q}_\parallel)]$. We first calculate $\partial_\omega{\rm Re}[ \chi_{\rm eff}(\vect{q}_\parallel,\omega)|_{\omega=\Omega(\vect{q}_\parallel)}]$ at leading order in $q_\parallel$. Only ${\chi}^{(0)}_{\rm AA}$ and ${\chi}^{(0)}_{\rm BB}$ contribute to this quantity at leading order in $q_\parallel$. We find $\partial_\omega{\rm Re}[ \chi_{\rm eff}(\vect{q}_\parallel,\omega)|_{\omega=\Omega(\vect{q}_\parallel)}]\to - q_{\parallel} [\cos(\theta)b/2 + vk^2_{\rm F}/(3\Omega_{\theta})]/(\hbar \Omega^2_\theta \pi^2)$. We then calculate ${\rm Im} [\chi_{\rm eff}(\vect{q}_\parallel, \omega)]$. As detailed in Appendix~\ref{app:AA_response}, ${\rm Im} [\chi^{\rm eff}_{\rm AA}(\vect{q}_\parallel,\omega)]\to - q_\parallel b\cos(\theta) \delta(\omega - E_{\rm A}({\bm q}_\parallel)/\hbar)/(2 \hbar \pi)$. This result can be easily understood as the absorption spectrum of a one-dimensional system~\cite{Giuliani_and_Vignale}, whose 
role in our 
case is effectively played by the FA states. For the calculation of ${\rm Im}[\chi^{\rm eff}_{\rm BB}(\vect{q}_\parallel,\omega)]$, we use the well-known approximate relation~\cite{Garrido} $\chi_{\rm BB}^{\rm eff}  (z, z^\prime ,\vect{q}_\parallel,\omega) \simeq \Theta(-z) \Theta (-z^\prime)[ \chi^{({\rm h})}_{\rm BB}(z+z^\prime,\vect{q}_\parallel, \omega)+ \chi^{({\rm h})}_{\rm BB} (z-z^\prime,\vect{q}_\parallel,\omega)]$. Here, $\chi^{({\rm h})}_{\rm BB}(z,\vect{q}_\parallel, \omega) = \int dq_{z} \chi^{({\rm h})}_{\rm BB}(\vect{q}_{\rm 3D}, \omega)\exp(iq_z z)/(2\pi)$ is the Fourier transform of the homogenous response function $\chi^{({\rm h})}_{\rm BB}(\vect{q}_{\rm 3D}, \omega)$ of a bulk WSM~\cite{lv_ijmp_2013,zhou_prb_2015} and ${\bm q}_{\rm 3D} = (q_z, {\bm q}_{\|})$. Straightforward algebraic manipulations yield
\begin{equation}\label{eq:ImBB}
\begin{split}
{\rm Im}[ \chi^{\rm eff}_{\rm BB}(\vect{q}_\parallel,\omega)]\simeq \int \frac{dq_z}{2\pi}  \frac{2q_\parallel^2}{\big(q_\parallel^2+q_z^2\big)^2} {\rm Im}[ \chi^{({\rm h})}_{\rm BB}(\bm{q}_{\rm 3D},\omega)]~.
\end{split}
\end{equation}
Eq.~(\ref{eq:ImBB}) is crucial as it states that, for every $\omega$, the decay rate (\ref{eq:LifeTime}) of the FA plasmon depends on the spectral density of electron-hole pairs ${\rm Im}[ \chi^{({\rm h})}_{\rm BB}(\bm{q}_{\rm 3D},\omega)]$ in the bulk integrated over all values of $q_{z}$. Physically, this equation expresses the fact that a surface plasmon can decay without conserving the $\hat{\bm z}$ component of the wave vector $\bm{q}_{\rm 3D}$, for the presence of the surface breaks translational invariance in this direction relaxing the three-dimensional momentum conservation that a bulk plasmon would obey. The latter is encoded in regions of the plane ${\bm q}_{\rm 3D}$-$\omega$ where ${\rm Im}[ \chi^{({\rm h})}_{\rm BB}(\bm{q}_{\rm 3D},\omega)]$ is zero. Because of the convolution in Eq.~(\ref{eq:ImBB}), this ceases to be true for ${\rm Im}[ \chi^{\rm eff}_{\rm BB}(\vect{q}_\parallel,\omega)]$.

In the limit $\omega \gg vq_\parallel$ and at the leading order in $2E_{\rm F}/(\hbar \omega) \gg 1$ we obtain
\begin{equation}\label{eq:ImBB2}
\begin{split}
{\rm Im} [\chi^{\rm eff}_{\rm BB}(\vect{q}_\parallel,\omega)]\to -\frac{q^2_{\|}}{32 \pi^2 \hbar \omega} \left(\frac{2E_{\rm F}}{\hbar\omega}\right)^2~.
\end{split}
\end{equation}
In the limit $k_{\rm F} \to 0$ we find ${\rm Im}[\chi^{\rm eff}_{\rm BB}(\vect{q}_\parallel,\omega)] = - q_{\|}/(24  \pi \hbar v )$. This implies that, in the undoped limit, the decay rate $\Gamma$ tends to a finite value in the long-wavelength $q_{\|} \to 0$ limit. The FA plasmon is therefore not a well defined excitation in the case of an undoped WSM, since it easily decays by emitting inter-band electron-hole pairs in the gapless bulk.

Finally, we need to calculate ${\rm Im}[\chi^{\rm eff}_{\rm AB}(\bm{q}_{\|},\omega)]$. Following the steps described in Appendix~\ref{app:AB_response}, we find
\begin{equation}\label{eq:ImAB}
\begin{split}
{\rm Im}[\chi^{\rm eff}_{\rm AB}(\vect{q}_\parallel,\omega)]\to -\frac{q_\parallel^2 \big[1+\sin^2(\theta)\big]}{3\pi^2 \hbar \omega}\left(\frac{2E_{\rm F}}{\hbar\omega}\right)^{-1/2}~.
\end{split}
\end{equation}
We emphasize that the contribution to the decay rate coming from Eq.~(\ref{eq:ImAB}) is suppressed with respect to that coming from Eq.~(\ref{eq:ImBB2}) by a factor $[2E_{\rm F}/(\hbar\omega)]^{-5/2}\ll 1$. We will therefore neglect the contribution (\ref{eq:ImAB}) for the calculation of $\Gamma$. Physically, Eq.~(\ref{eq:ImAB}) represents processes in which a FA plasmon decays by emitting a composite electron-hole pair, with one partner of the pair belonging to the FA manifold of states and the other partner to the bulk manifold. In the limit $k_{\rm F} \to 0$, we get ${\rm Im} [ \chi^{\rm eff}_{\rm AB}(\bm{q}_{\|},\omega)] = - 2 q^2_{\|} [1+\sin^2(\theta)]/(3  \pi^2 \hbar \omega)$.

In summary, for a doped WSM we get the following compact expression for the decay rate of a FA plasmon in the long-wavelength limit and deep in the single-particle optical gap $\hbar \omega \ll 2E_{\rm F}$:
\begin{equation}\label{eq:Q}
\begin{split}
\Gamma(\vect{q}_\parallel)= \frac{\Omega_\theta}{32} \frac{q_\parallel}{k_{\rm F}} \frac{\displaystyle 1}{\displaystyle \frac{\cos{(\theta)}}{2} \frac{b}{k_{\rm F}} + \frac{1}{6}\left(\frac{2 E_{\rm F}}{\hbar \Omega_\theta}\right)}\left(\frac{2E_{\rm F}}{\hbar\Omega_\theta}\right)^2~.
\end{split}
\end{equation}
The quantity $\Gamma(\vect{q}_\parallel)/\Omega(\vect{q}_\parallel)$ is plotted in Figs.~\ref{fig:two}(b) and~\ref{fig:three}(b). In writing Eq.~(\ref{eq:Q}) we have omitted the contribution $\propto \delta(\omega - E_{\rm A}({\bm q}_\parallel)/\hbar)$ stemming from ${\rm Im}[\chi^{\rm eff}_{\rm AA}]$. This $\delta$-function contribution is not present in the range of values of ${\bm q}_{\|}$ we have used to make the plot in Fig.~\ref{fig:two}(b). From Fig.~\ref{fig:two}(b) we immediately see that the WSM surface plasmons are highly directional and weakly damped only in the direction of propagation of the single-particle FA states. From Figs.~\ref{fig:two}(b)-\ref{fig:three}(b) we clearly see that damping increases (quality factor decreases) very rapidly as a function of the angle $\theta$.

Before concluding, we would like to comment on one evident limitation of the WSM model in Eq.~(\ref{eq:hamiltonian}). As discussed above, in this model the FA states disperse only in the $\hat{\bm k}_{y}$ direction and their dispersion is strictly linear, $E_{\rm A}({\bm k}_{\|}) = \hbar v k_{y}$. More complicated models---see, for example, Ref.~\onlinecite{tchoumakov_prb_2017}---allow of course for more general FA dispersion relations. For example, taking $E_{\rm A}({\bm k}_{\|})=\hbar v k_{y}+ \hbar^2(k_x^2-b^2)/(2 m_{x})$, one can demonstrate that Eq.~(\ref{eq:OmegaP}) is not modified by the parabolic correction. Also, one can show that the first non-local correction in Eq.~(\ref{eq:OmegaP1}) is modified only for $\theta \sim \pm \pi/2$, i.e.~along the directions parallel or anti-parallel to the Weyl-node separation vector $2{\bm b}$. Along these directions, however, the FA plasmon is anyway strongly damped---see Fig.~\ref{fig:three}(b)---and therefore quantitative changes to its dispersion 
relation for $\theta \sim \pm \pi/2$ are uninteresting.

\begin{figure}[t]
\centering
\begin{overpic}[width=0.8\columnwidth]{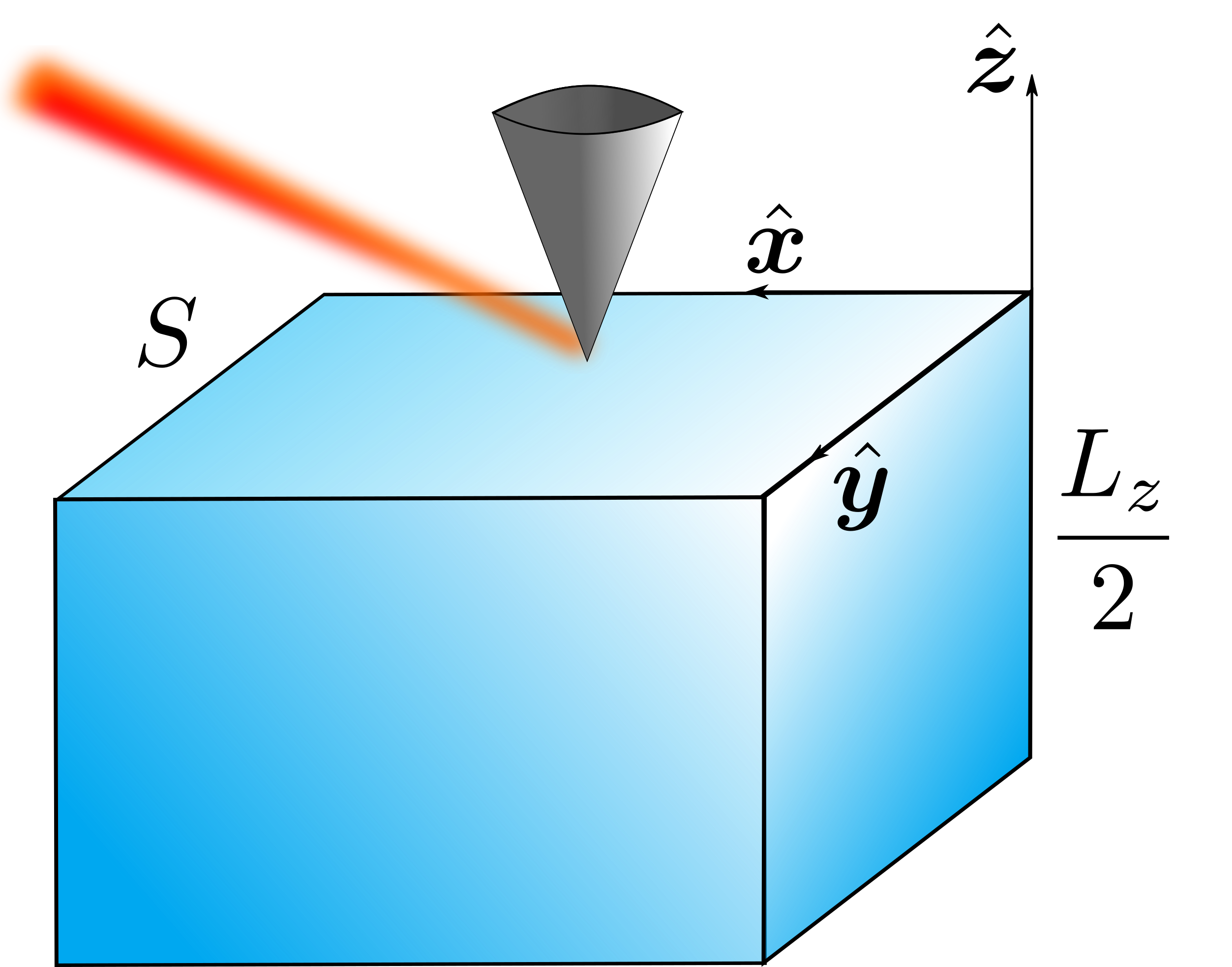}\put(-10,65){\normalsize (a)}\end{overpic}\vspace{0.5em}
\begin{overpic}[width=1\columnwidth]{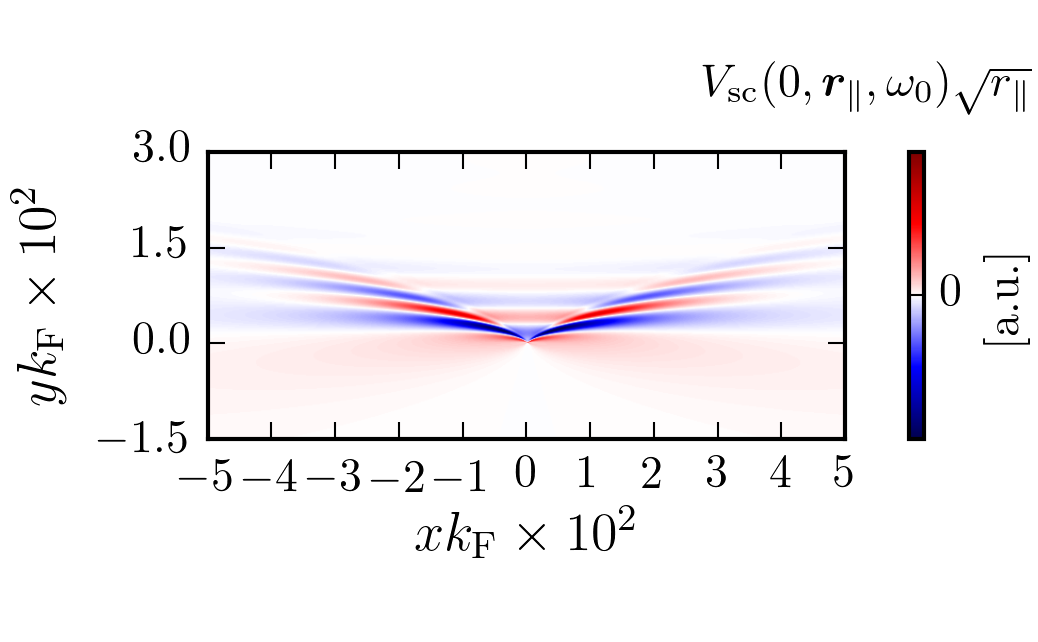}\put(2,52){\normalsize (b)}\end{overpic}
\caption{(Color online)  Panel (a) shows the interface, located on the $\hat{\bm x}$-$\hat{\bm y}$  plane, between a semi-infinite WSM and vacuum. The WSM sample (blue) is a parallelepiped with surface area $S$ and height $L_{z}/2$. The (grey) cone represents the metallized tip of an atomic force microscope used in a s-SNOM experiment. The latter is illuminated by a laser (red) and located at the origin of the $\hat{\bm x}$-$\hat{\bm y}$ plane. In panel (b) we present the screened potential in response to a laser with frequency $\omega_0$.
The color map shows the calculated screened potential evaluated at $z=0$ and multiplied by the square-root of the distance $r_\parallel$ from the tip located at the origin, i.e.~$V_{\rm sc}(0, \bm{r}_\parallel, \omega) \sqrt{r_\parallel}$. 
Results shown in this panel have been obtained by setting $b/k_{\rm F}=5$, $\alpha_{\rm ee}=0.8$, and $\hbar \omega/E_{\rm F}=1.4$.\label{fig:setup}}
\end{figure}
\section{Response of FA plasmons to an illuminated metallized tip}
\label{sect:SNOM}

In this Section we present an elementary theory that describes the potential that one would observe by carrying out a scattering-type near-field optical experiment (s-SNOM)~\cite{basov_science_2016} on the surface of a WSM with BTRS. In this experiment~\cite{basov_science_2016}, light is focused on a metallic tip with the aim of launching propagating surface plasmons. In Fig.~\ref{fig:setup}(a) we sketch the setup of such an experiment.
 
We model the tip as an external potential oscillating in time and with an exponential profile in the $\hat{\bm z}$ direction:
\begin{equation}
v_{\rm ext}(z,\vect{q}_\parallel,t)=V_0e^{-q_\parallel |z|}e^{-i\omega_{0} t}~,
\end{equation}
where $\omega_{0}$ is the frequency of the laser that illuminates the tip. The exponential dependence on $z$ dramatically simplifies the algebra. In the Fourier transform, we find $v_{\rm ext}(z,\vect{q}_\parallel,\omega) \sim \delta(\omega - \omega_{0})$. Since we are interested in what happens at the surface of a WSM, we set $z=0$. In this case, the external potential reduces to a delta function of $\vect{r_\parallel}$: $v_{\rm ext}(0,\vect{r}_\parallel,t) = V_0\delta(\vect{r}_\parallel)e^{-i\omega_{0} t}$.

Working in the RPA, we can easily write the relationship between $V_{\rm sc}(0,\vect{q}_\parallel,\omega)$ and the external potential $v_{\rm ext}(0,\vect{q}_\parallel,\omega)$, which involves the effective 2D response function $\chi_{\rm eff}(\vect{q}_\parallel,\omega_{0})$:
\begin{equation}
  V_{\rm sc}(0,\vect{q}_\parallel,\omega_{0})=\frac{V_0}{\displaystyle 1-\frac{2\pi e^2}{q_	\parallel} \chi_{\rm eff}(\vect{q}_\parallel,\omega_{0}) } ~.
\end{equation}
Near resonance, i.e.~when the laser frequency $\omega_{0}$ is close to the WSM FA plasmon frequency $\Omega(\vect{q}_\parallel)$, we can write
\begin{widetext}
\begin{equation}
  V_{\rm sc}(0,\vect{q}_\parallel,\omega_{0})\simeq \frac{V_{0}}{\displaystyle - \frac{2\pi e^2}{q_\parallel} \left(\frac{\partial \chi_{\rm eff}(\vect{q}_\parallel,\omega) }{\partial \omega}\right)_{\omega=\Omega(\vect{q}_\parallel)} \big[\omega_{0}-\Omega(\vect{q}_\parallel)+i\Gamma(\vect{q}_\parallel)\big ] }~.
\end{equation}
%
Fourier-transforming the previous result to real space we find 
%
%
\begin{equation}
V_{\rm sc}(0,\vect{r}_\parallel,\omega_{0})\simeq  V_0 \int \frac{d^{2}\vect{q}}{(2\pi)^2}\frac{e^{i\vect{q}_\parallel\vect{r}_\parallel}}{\displaystyle -\frac{2\pi e^2}{q_{\parallel}} \left(\frac{\partial \chi_{\rm eff}(\vect{q}_\parallel,\omega) }{\partial \omega}\right)_{\omega=\Omega(\vect{q}_\parallel)} \big[\omega_{0}-\Omega(\vect{q}_\parallel)+i\Gamma(\vect{q}_\parallel)\big ] }~.
\end{equation}
%
This integral can be easily evaluated in polar coordinates:
%
%
\begin{equation}
 V_{\rm sc}(0,\vect{r}_\parallel,\omega_{\rm TIP})\simeq V_{0} 
 \int_{0}^{\infty} 
 q_{\parallel} dq_{\parallel} \int_{0}^{2\pi} d\theta \frac{1}{(2\pi)^2}
 \frac{e^{iq_\parallel r_\parallel \cos(\theta-\theta^\prime)}}{\displaystyle -\frac{2\pi e^2}{q_\parallel} 
 \left(\frac{\partial \chi_{\rm eff}(q_\parallel,\theta,\omega) }{\partial \omega}\right)_{\omega=\Omega(\vect{q}_\parallel)} 
 \big[\omega_{0}-\Omega(q_\parallel,\theta)+i\Gamma(q_\parallel,\theta)\big ] }~,
\end{equation}
\end{widetext}
where we have introduced the angle $\theta^{\prime}$ defined by ${\vect{r}_\parallel \cdot \vect{\hat{y}}}=r_\parallel \cos(\theta^\prime)$.

Using the result $\partial_\omega{\rm Re}[ \chi_{\rm eff}(\vect{q}_\parallel,\omega)|_{\omega=\Omega(\vect{q}_\parallel)}]\to - q_{\parallel} [\cos(\theta)b/2 + vk^2_{\rm F}/(3\Omega_{\theta})]/(\hbar \Omega^2_\theta \pi^2)$ reported above, and noting that the integrand is peaked where the denominator vanishes, we can extend the integration range over $q_{\parallel}$ from $[0,\infty]$ to $[-\infty,\infty]$. This allows us to use the power of the residue theorem.

The integrand has a simple pole with respect to $q_{\parallel}$. We define it as $\tilde{p}(\omega_{0},\theta)=p_1+i p_2$, which satisfies the equation 
$\omega_{0}=\Omega(\tilde{p},\theta)-i\Gamma(\tilde{p},\theta)$. Using the definitions of $\mathcal{I}(\theta)$ and $\Gamma_\theta$ given above, we have
\begin{equation}
p_1 = \frac{\omega_{0}-\Omega_\theta}{\Gamma_\theta^2 + [v \alpha \mathcal{I}(\theta)]^2} v \alpha\mathcal{I}(\theta)
\end{equation}
and
\begin{equation}
p_2= \frac{\omega_{0} - \Omega_\theta}{\Gamma_\theta^2+[v \alpha\mathcal{I}(\theta)]^2} \Gamma_\theta~.
\end{equation}
After straightforward mathematical manipulations we find
\begin{eqnarray}
 V_{\rm sc}(0,\vect{r}_\parallel,\omega_{0}) &\simeq& V_{0} \int_{\theta^\prime - \pi/2}^{\theta^\prime + \pi/2} d\theta \frac{\tilde{p}(\omega_{0},\theta) \Omega_\theta^2}{\cos(\theta)b/2 + vk^2_{\rm F}/(3\Omega_{\theta})}  \nonumber \\ 
 &\times & \exp{[i \tilde{p}(\omega_{0},\theta)r_{\parallel} \cos(\theta-\theta^\prime)]}~,
\end{eqnarray}
which needs to be evaluated numerically.

Fig.~\ref{fig:setup}(b) displays the calculated screened  potential for the following parameters: $b/k_{\rm F}=5$, $\alpha_{\rm ee}=0.8$, and $\hbar \omega/E_{\rm F}=1.4$. 
Here, it is possible to recognize features due to the peculiar propagation dynamics of FA plasmons. 
The illustrated SNOM pattern is highly anisotropic because of the highly unidirectional character of both FA dispersion and damping. 
The anisotropy of the dispersion relation---shown in Fig.~\ref{fig:two}(a)---manifests through the FA plasmon group velocity, which is maximal around the angular directions $\theta \approx \pm \pi/4$.
The anisotropy of the plasmon dissipation---see Fig.~\ref{fig:two}(b)---shows up in the propagation direction of FA plasmons, which predominantly occurs only along the $y>0$ direction.

\section{Summary and conclusions}
\label{sect:summary}

In summary, we have presented a quantum-mechanical non-local theory of surface plasmons in semi-infinite Weyl semimetals with broken time-reversal symmetry. We have been able to derive a simple analytical formula---see Eqs.~(\ref{eq:OmegaP})-(\ref{eq:I})---for the surface plasmon dispersion relation in the electrostatic limit $|{\bm q}_{\|}|\gg \omega/c$, which takes into account exquisite quantum effects associated with the penetration of the Fermi arc surface states into the gapless bulk. We have also included non-local corrections, which were crucial to investigate in a quantitative manner the surface plasmon damping rate (\ref{eq:Q}), as determined by decay processes involving the excitation of electron-hole pairs. 

Our calculations show that the intrinsic damping of topological Fermi arc plasmons is small at small values of the in-plane wave vector ${\bm q}_{\|}$, mainly in a specific directions, i.e.~the direction along which the single-particle Fermi arc states disperse. Scattering-type near-field 
optical spectroscopy can therefore be used as an alternative to ARPES to carry out spatially-resolved investigations of these intriguing chiral modes occurring in Weyl semimetals with broken time-reversal symmetry, in the absence of an external magnetic field.

\acknowledgments
It is a great pleasure to thank Iacopo Torre and Fabio Taddei for very useful discussions. M.P. wishes to thank Fondazione Istituto Italiano di Tecnologia for financial support. 
F.H.L.K. acknowledges financial support from the Government of Catalonia through the SGR grant (2014-SGR-1535), and from the Spanish Ministry of Economy and Competitiveness, through the "Severo Ochoa" Programme for Centres of Excellence in R\&D (SEV-2015-0522), support by Fundacio Cellex Barcelona, CERCA Programme / Generalitat de Catalunya and the Mineco grants Ram\'{o}n y Cajal (RYC-2012-12281) and Plan Nacional (FIS2013-47161-P and FIS2014-59639-JIN). Furthermore, the research leading to these results has received funding from the European Union's Horizon 2020 research and innovation programme under grant agreement No.~696656 ``GrapheneCore1'' and the ERC starting grant (307806, CarbonLight). 
M.P. is extremely grateful for the financial support granted by ICFO during a visit in August 2016. 

\appendix

\section{Boundary conditions}
\label{app:BCs}
In order to determine the correct boundary conditions, we solve the problem for $z>0$, evaluating the evanescent states at $z=0^{+}$, and then we take the $\lim_{\tilde{m}\to +\infty}{\cal H}_{m(z)}({\bm k}_{\|}, k_{z}\to -i\partial_z)$ in the semi-infinite WSM model Hamiltonian introduced in the main text.
We find
\begin{equation}\label{eq:Boundary}
\begin{split}
\lim_{\tilde{m}\rightarrow+\infty} \Psi_{E,\vect{k}_\parallel} (\vect{r}_\parallel,0^+) \propto
\begin{pmatrix}
1\\
0
\end{pmatrix} e^{i\vect{k}_\parallel\vect{r}_\parallel}~,
\end{split}
\end{equation}
where $\Psi_{E,\vect{k}_\parallel}(\vect{r}_\parallel,z)$ is the evanescent solution of the problem for $z>0$. Imposing the continuity of the wave function at $z=0$ between the reflected state in Eq.~(4) of the main text and the evanescent state~(\ref{eq:Boundary}) we find $r_{\vect{k}}=-1$. Note that the square modulus of the reflected state in Eq.~(4) is correctly normalized only in the limit $V\rightarrow \infty$.

\section{Dielectric background screening}
\label{app:screening}

We here describe how to include the effect of a dielectric background in Eq.~(7) of the main text.

We assume that dielectric screening stems from the response of the empty (occupied) states in the bands above (below) the conduction (valence) band, far away in energy from the Weyl crossing.

The total response $\chi_{\rm tot}$ of the system is the sum of two terms: i) the response of the states described by the Weyl Hamiltonian in Eq.~(1) of the main text, denoted by the symbol $\chi$, and ii) the contribution due to the high-energy bands, denoted by the symbol $\chi_{\rm b}$, i.e.
\begin{equation}\label{eq:resp_wbg}
\chi_{\rm tot}\equiv \chi+\chi_{\rm b}~.
\end{equation}
For simplicity,  we start from the homogeneous (h) case and we focus on a single pair of bands, with one energetically higher than the conduction band 
and the other one energetically lower than the valence band. By using the Lehmann representation we can write
\begin{eqnarray}\label{eq:bg}
\chi^{(\rm h)}_{\rm b}(\vect{q}_{\rm 3D},\omega)&=& \frac{1}{V} \sum_{{\rm L}, {\rm H}} \frac{f_{\rm L}-f_{\rm H}}
 {\epsilon_{\rm L}-\epsilon_{\rm H}+\hbar\omega}
  |\mathcal{M}_{{\rm L}, {\rm H}} (\vect{q}_{\rm 3D})|^2 \nonumber \\ &+&
  \big[\omega\rightarrow-\omega,\vect{q}_{\rm 3D}\rightarrow-\vect{q}_{\rm 3D}\big]\,, 
\end{eqnarray}
where the collective labels ``L'' and ``H'' refer to states of a given electronic band with lower (higher) energy than the valence (conduction) band, respectively, and
${\cal M}_{{\rm L}, {\rm H}}  (\vect{q}_{\rm 3D})$ represents a suitable matrix element. 
Because the band L (H) is deep below (well above) the Fermi level, we replace $f_{\rm L} \to 1$ ($f_{\rm H} \to 0$).
Furthermore, we define $\Delta\equiv\min(\epsilon_{\rm H} - \epsilon_{\rm L}) > 0$. Since we are interested in the frequency range $\omega \ll \Delta/\hbar$ we can approximate Eq.~(\ref{eq:bg}) as

\begin{equation}\label{eq:bg_omega}
\chi^{(\rm h)}_{\rm b}(\vect{q}_{\rm 3D},\omega)\simeq -\frac{1}{ \Delta V} \sum_{{\rm L}, {\rm H}}[|{\cal M}_{{\rm L}, {\rm H}} ({\bm q}_{\rm 3D})|^2+|\mathcal{M}_{{\rm L}, {\rm H}} (-\vect{q}_{\rm 3D})|^2]~. 
\end{equation}

In the long-wavelength limit, because of the orthogonality of the states belonging to different bands, we have 
$  |\mathcal{M}_{{\rm L}, {\rm H}}  (\vect{q}_{\rm 3D})|^2 \propto \vect{q}_{\rm 3D}^2$. We can therefore write 
\begin{equation}\label{eq:bg_q}
\chi^{(\rm h)}_{\rm b}(\vect{q}_{\rm 3D},\omega)\simeq -\chi_0\vect{q}_{\rm 3D}^2\,,
\end{equation}
where $\chi_{0}$ is a positive constant, independent of $\vect{q}$ and $\omega$. 
By exploiting the generality of the above argument, we can express the global contribution to the response function from all electronic bands far away in energy from the Weyl crossing as $\chi^{(\rm h)}_{\rm b}(\vect{q}_{\rm 3D},\omega) \simeq -\chi_{\rm b} \vect{q}_{\rm 3D}^2$,
where $\chi_{\rm b} >0$.

In order to recover the plasmon dispersion relation in the homogeneous case, we have to solve the well-known RPA equation~\cite{Giuliani_and_Vignale}
\begin{equation}\label{eq:RPAbulk}
\frac{4\pi e^2}{\vect{q}_{\rm 3D}^2} \chi^{(\rm h)}_{\rm tot}(\vect{q}_{\rm 3D},\omega)=1~,
\end{equation}
or, equivalently,
\begin{equation}\label{eq:epsb_bulk}
 \frac{4\pi e^2}{\epsilon_{\rm b} \vect{q}_{\rm 3D}^2} \chi^{(\rm h)}(\vect{q}_{\rm 3D},\omega)=1\,,
\end{equation}
where $\epsilon_{\rm b}=1+ 4 \pi e^2 \chi_{\rm b}$.

We now follow a similar path for the case of surface plasmon modes.
We first replace the expression $\chi_{\rm b}  (z, z^\prime ,\vect{q}_\parallel,\omega)  
\simeq\Theta(-z) \Theta (-z^\prime)[ \chi^{(\rm h)}_{\rm b}(z+z^\prime,\vect{q}_\parallel)+ \chi^{(\rm h)}_{\rm b} (z-z^\prime,\vect{q}_\parallel,\omega)]$ in Eq.~(\ref{eq:resp_wbg}). We then use the result derived earlier, i.e.~$\chi^{(\rm h)}_{\rm b}(\vect{q}_{\rm 3D},\omega) \simeq -\chi_{\rm b} \vect{q}_{\rm 3D}^2$. Carrying out straightforward algebraic manipulations we find Eq.~(7) in the main text, i.e.~
\begin{equation}\label{eq:RPA_}
1=\frac{2\pi e^2}{\bar{\epsilon}q_\parallel}\chi_{\rm eff}(\vect{q}_\parallel,\omega)~,
\end{equation}
where $\bar{\epsilon} = (1+\epsilon_{\rm b})/2$.

\section{The AA response function}
\label{app:AA_response}

We here provide more details on the calculation of the response function ${\chi}^{(0)}_{\rm AA}(q_z,q_z^{\prime},\vect{q}_\parallel,\omega)$. 
Using the FA eigenstates in Eq.~(3) of the main text and the Lehmann representation~\cite{Giuliani_and_Vignale}---Eq.~(10) of the main text---we find the following exact expression:
\begin{eqnarray}
\label{eq:AA}
{\chi}^{(0)}_{\rm AA}(q_z,q_z^{\prime},q_x,q_y,\omega)=\frac{1}{(2\pi)^2L_z} \frac{q_y}{\hbar(\omega+i 0^+)-vq_y} \nonumber  \\  \times \int_{-b}^{b-|q_x|} dk_x \mathcal{L}(q_z,q_z^{\prime},q_x)~, \nonumber  \\
\end{eqnarray}
where
\begin{eqnarray}
\mathcal{L}(q_z,q_z^{\prime},q_x)&\equiv &\left \{\frac{2\big[b^2-(k_x+|q_x|)^2\big]}{\big(b^2-k_x^2\big)+\big[b^2-(k_x+|q_x|)^2\big]-i2bq_z} \right \} \nonumber \\ &\times & \left \{
 \frac{2 \big(b^2-k_x^2\big)}{\big(b^2-k_x^2\big)+\big[b^2-(k_x+|q_x|)^2\big]+i2bq^\prime_z} \right \}~. \nonumber \\ 
\end{eqnarray}
Because in our model the FA states do not disperse along the $\hat{\bm x}$ direction, 
we have ${\chi}^{(0)}_{\rm AA}(q_z,q_z^{\prime},q_{x},q_{y}=0,\omega)=0$. This means that, in our model, an external field can perturb the FA states only if it carries a finite momentum along the  $\hat{\bm y}$ direction. Expanding $\mathcal{L}(q_z,q_z^{\prime},q_x)$ in a power series of $q$ and carrying out the integral in Eq.~(\ref{eq:AA}), we obtain 
\begin{equation}\label{eq:REAA}
{\rm Re}[{\chi}^{(0)}_{\rm AA}(q_z,q_z^{\prime},\vect{q}_\parallel,\omega)]\to\frac{1}{(2\pi)^2L_z} \frac{q_y}{\hbar \omega} \left[ 2b\Big(1+\frac{vqy}{\omega}\Big)-|q_x|\right]~.
\end{equation}
From Eq.~(\ref{eq:REAA}) one easily obtains ${\rm Re}[{\chi}^{\rm eff}_{\rm AA}({\bm q}_{\|},\omega)]$, as in Eq.~(12) of the main text.
Similarly, we find
\begin{equation}\label{eq:ImAA}
{\rm Im}[{\chi}^{(0)}_{\rm AA}(q_z,q_z^{\prime},\vect{q}_\parallel,\omega)]\to-\frac{q_y}{4\pi L_z} \big( 2b -|q_x|  \big) \delta(\omega-vq_y)~,
\end{equation}
from which ${\rm Im}[{\chi}^{\rm eff}_{\rm AA}({\bm q}_{\|},\omega)]$ follows.

\section{The BB response function}
\label{app:BB_response}

We here report some useful technical details on the calculation of the response function ${\chi}^{(0)}_{\rm BB}(q_z,q_z^{\prime},\vect{q}_\parallel,\omega)$ for the doped case.

Since we are interested in studying the long-wavelength limit, we can use the (high-frequency) moment expansion~\cite{Giuliani_and_Vignale,torre_prb_2017}, i.e.
\begin{equation}\label{eq:ChiBB_SOM}
 {\rm Re}[ {\chi}^{(0)}_{\rm BB}](q_z,q_z^{\prime},\vect{q}_\parallel,\omega)\to 
 \frac{\braket{\hat{n}_{q_z-q_z^\prime}} _{\rm GS} ({q_\parallel^2+q_zq_z^\prime})}{m_{\rm eff} \omega^2 V}\,,
\end{equation}
where $m_{\rm eff}=E_{\rm F}/v^2$ is the effective mass, $\hat{n}_{q_{z} - q^\prime_{z}}=\sum_j e^{i (q_z-q_z^\prime)\hat{z}_j}$ is the Fourier Transform of the density operator~\cite{Giuliani_and_Vignale}, 
and $\braket{\ldots}_{\rm GS}$ is a shorthand for the following ground-state expectation value $\bra{\rm GS}_{\rm B}\ldots \ket{\rm GS}_{\rm B}$. Here, $\ket{{\rm GS}}_{\rm B}=\prod_{|{\bm k}|<k_{\rm F}} b_{+,{\bm k}}^\dagger \ket{0}$, where $ b_{+1,{\bm k}}^\dagger$ creates an electron in a state labeled by the band index $\lambda=+1$ 
and wavevector ${\bm k}$. The quantity $\ket{0}$ represents the vacuum state with no electrons in the conduction band.

By using the eigenstates in Eq.~(4) of the main text, we find that
\begin{equation}\label{Eq:densityBB}
\braket{\hat{n}_{q_{z}}}_{\rm GS}=\braket{\hat{n}_{q_{z}}}_{\rm GS}^{({\rm scl})}+\braket{\hat{n}_{q_{z}}}_{\rm GS}^{({\rm int})}
\end{equation}
where
\begin{eqnarray}\label{Eq:density_scl}
\braket{\hat{n}_{q_z}}_{\rm GS}^{({\rm scl})} &\equiv & 2 S \int_{k<k_{\rm F}, k_{z}>0} \frac{d^{3}\vect{k}}{(2\pi)^3} \int dz  \nonumber \\ &\times & e^{-i q_z z } \Theta(-z) 
\Big[|	u_{{\bm k},+1}|^2+|	u_{\bar{\bm k},+1}|^2\Big]
\end{eqnarray}
and
\begin{eqnarray}\label{Eq:density_int}
 &\braket{\hat{n}_{q_{z}}}_{\rm GS}^{({\rm int})}\equiv-2 S \int_{k<k_{\rm F}, k_{z}>0} \frac{d^{3}\vect{k}}{(2\pi)^3} \int dz   \nonumber \\& \times  e^{-i q_z z } \Theta(-z) 
\Big[u_{\bar{{\bm k}},+1}^\dagger u_{{\bm k},+1}e^{i2k_z z}+u^\dagger_{{\bm k},+1}u_{\bar{{\bm k}},+1}e^{-i2k_z z}\Big]~,\nonumber \\
\end{eqnarray}
where $u_{{\bm k}, \lambda=\pm 1}$ has been introduced in Eq.~(2) of the main text.
As in the main text, $\bar{\bm k}=({\bm k}_{\|},-k_z)$.
The index ``scl'' (``int'') stands for ``semiclassical'' (``interference''). Indeed,  $\braket{\hat{n}_{q_z}}_{\rm GS}^{({\rm scl})}$ ($\braket{\hat{n}_{q_{z}}}_{\rm GS}^{({\rm int})}$) is independent of (dependent of) the relative phases, see Eq.~(4) of the main text. 

We start from the case $q_{z} =0$. We find
\begin{eqnarray}
\braket{\hat{n}_{0}}_{\rm GS}=N+\braket{\hat{n}_{0}}^{({\rm int})}_{\rm GS}~,
\end{eqnarray}
where $N$ is the total number of electrons in the semi-infinite WSM,
while the interference term $\braket{\hat{n}_{0}}_{\rm GS}^{\rm int} \propto 1/L_z$ is negligible in the thermodynamic limit. 

For a generic value of $q_{z}$, we obtain
\begin{eqnarray}\label{eq:scl_qz}
\braket{\hat{n}_{q_z}}_{\rm GS}^{({\rm scl})}=n S \left[-\frac{1}{iq_z}+\pi \delta({q_z})\right]~,
\end{eqnarray}
where $n \equiv 2N/(L_z S)$. We then calculate the interference term $\braket{\hat{n}_{q_z}}^{({\rm int})}_{\rm GS}$.
We are interested in the small $q_{z}$ limit. We find $\braket{\hat{n}_{q_z}}^{({\rm int})}_{\rm GS} - \braket{\hat{n}_{0}}^{({\rm int})}_{\rm GS} \propto q_{z}$. In the thermodynamic limit, this can induce in ${\rm Re}[{\chi}^{(0)}_{\rm BB}(q_z,q_z^{\prime},\vect{q}_\parallel,\omega)]$ terms of ${\cal O}(q^{3}_{z}, q^{\prime3}_{z})$, which are beyond the interest of our leading-order long-wavelength theory.

Finally, taking into account only the semiclassical term in Eq.~(\ref{eq:ChiBB_SOM}), we find
\begin{eqnarray}
\label{eq:ChiBB}
{\rm Re}[{\chi}^{(0)}_{\rm BB}(q_z,q_z^{\prime},\vect{q}_\parallel,\omega)]=\frac{n\big(q_\parallel^2+q_zq_z^\prime\big)}{m_{\rm eff} L_z\omega^2} \nonumber \\ \times\left[\frac{1}{i(q^\prime_z-q_z)}+\pi \delta(q_z-q_z^\prime)\right]~, \nonumber \\ 
\end{eqnarray}
which implies Eq.~(14) of the main text.

\section{The AB response function}
\label{app:AB_response}

Here, we detail the calculation of the response function $\chi^{(0)}_{\rm AB}$, always for a generic doping. We neglect inter-Weyl-node electron-electron scattering processes and we linearize the spectrum $E_{\lambda}({\bm k})$ around each Weyl node. In order to simplify the notation we set $\hbar=1$. Under these simplifying assumptions, the quantity we need to calculate is
\begin{widetext}
\begin{eqnarray}\label{eq:chiAB_explicit}
\chi^{(0)}_{\rm AB}(q_z,q_z^\prime,\vect{q}_\parallel,\omega)=\sum_{j={\pm 1},\lambda={\pm 1}}\int  \frac{d^{3}\vect{k}}{L_z(2\pi)^3}\Theta(-j k_x+j q_x) \Theta(k_z)\frac{\Theta(k_F+q_y-k_y)-\Theta(k_\lambda-k)}{vk_y-vq_y-\lambda vk+\omega+i\eta} \mathcal{M}^j_\lambda(q_x,q_z,q_z^\prime)
\nonumber\\+
\sum_{j={\pm 1},\lambda={\pm 1}}\int  \frac{d^{3}\vect{k}}{L_z(2\pi)^3}\Theta(-j k_x-j q_x) \Theta(k_z)\frac{\Theta(k_F-q_y-k_y)-\Theta(k_\lambda-k)}{vk_y+vq_y-\lambda vk-\omega-i\eta} \mathcal{M}^j_\lambda(-q_x,-q_z,-q_z^\prime)~,
\end{eqnarray}
\end{widetext}
where $\eta = 0^{+}$, $j=\pm 1$ is a Weyl node index, $\lambda=\pm$ is a conduction/valence band index, $k_{+1}=k_{\rm F}$, and $k_{-1}=\Lambda$. Here, $\Lambda$ is an ultraviolet cutoff. We will take the limit $\Lambda \to \infty$ momentarily. The matrix element $ \mathcal{M}^j_\lambda(q_x,q_z,q_z^\prime)$ reads as follows
\begin{widetext}
\begin{eqnarray}
\mathcal{M}^j_\lambda(q_x,q_z,q_z^\prime)&=&\frac{-j( k_x-q_x)}{L_z} [1+\lambda\cos(\beta_{k})] \frac{1}{k_x^2+k_z^2}
 \biggl[\frac{j k_x- i k_z}{-j (k_x-q_x)+i(k_z-q_z)} - \frac{j k_x+ik_z  }{-j (k_x-q_x)-i(k_z+q_z)}\biggr]\nonumber  \\ 
 &\times& \biggl[\frac{j k_x+ik_z  }{-j (k_x-q_x)-i(k_z-q^\prime_z)}-\frac{j k_x-ik_z  }{-j (k_x-q_x)+i(k_z+q^\prime_z)}\biggr]~.
\end{eqnarray}
\end{widetext}
We denote by the shorthand $q$ the small quantity $q_z \sim q_z^\prime \sim q_{x}$, and we expand the response function up to the quadratic order in $q$. Carrying out the sum over the index $j$ we find
\begin{widetext}
\begin{eqnarray}
\chi^{(0)}_{\rm AB}(q_z,q_z^\prime,\vect{q}_\parallel,\omega)=-\frac{{ q_x^2+q_z q_z^\prime }}{L_z\pi^3}\sum_{\lambda}\int  d^{3}\vect{k}\bigg\{\frac{\Theta( k_x) \Theta(k_z)[\Theta(k_{\rm F}-k_y)-\Theta(k_\lambda -  k)]}{v^2(k_y-\lambda k)^2-(\omega+i\eta)^2} \bigg\}\frac{ 2\lambda v k_x k_z^2}{k\big(k_x^2+k_z^2\big)^2}~.  
\end{eqnarray}
\end{widetext}

It is now useful to define the following dimensionless quantities, $\vect{p}\equiv{\vect{k}}/{k_{\rm F}}$ and $\tilde{\omega}\equiv{\omega}/{E_{\rm F}}$, and use cylindrical coordinates, $p_z=p_\perp \cos(\phi)$ and $p_y=p_\perp \sin(\phi)$. 
We perform the integral over $\phi$ and, exploiting the identity  $\lim_{\Lambda\rightarrow \infty} \big[\Theta(1-p_y)-\Theta(\Lambda/k_{\rm F}-p)\big]=-\Theta(p_y-1)$, we obtain
\begin{eqnarray}\label{eq:chi0AB_SOM}
\chi^{(0)}_{\rm AB}(q_z,q_z^\prime,\vect{q}_\parallel,\omega)=\frac{{ q_x^2+q_z q_z^\prime }}{3L_z\pi^3E_{\rm F}}
\left[\mathcal{J}_1(\omega)-\mathcal{J}_2(\omega)-\mathcal{J}_3(\omega)\right]~,\nonumber \\
\end{eqnarray}
where we have defined
\begin{widetext}
\begin{eqnarray}
\mathcal{J}_1(\omega)&=&\int_{-\infty}^{+\infty}  d p_y \int _0^\infty dp_\perp \frac{\Theta(1 - p)}{ p}\bigg[\frac{2}{(p_y- p)^2-(\tilde{\omega}+i\eta)^2} \bigg]\,,   \\
\mathcal{J}_2(\omega)&=&\int_{-\infty}^{+\infty}  d p_y \int _0^\infty dp_\perp  \frac{\Theta(1 - p_y)}{ p}\bigg[\frac{2}{(p_y-p)^2-(\tilde{\omega}+i\eta)^2} \bigg]\,,  \\
\mathcal{J}_3(\omega)&=&\int_{-\infty}^{+\infty}  d p_y \int_0^\infty dp_\perp\frac{ \Theta(p_y-1) }{ p}\bigg[\frac{2}{(p_y+ p)^2-(\tilde{\omega}+i\eta)^2} \bigg]~.  
\end{eqnarray}
\end{widetext}
We first analyze the real part of the response function. We start by studying the auxiliary function $\mathcal{J}_1(\omega)$. 
Introducing a new set of polar coordinates, $p_y=p \sin (\theta)$ and $p_\perp =p\cos (\theta)$, we find 
\begin{eqnarray}
{\rm Re}[\mathcal{J}_1(\omega)]={\rm PV}\int _{-\pi/2}^{+\pi/2}d \theta \int _0^1 dp \frac{2}{p^2[\sin(\theta)- 1]^2-\tilde{\omega}^2}~,    \nonumber \\
\end{eqnarray}	
where ``${\rm PV}$'' denotes the Cauchy principal value.
The following mathematical identity will be used below:
\begin{eqnarray}
 {\rm PV} \int  dp \frac{1}{p^2[1-\sin(\theta)]^2 - \tilde{\omega}^2} = \nonumber \\ \frac{1}{2\tilde{\omega} [1-\sin(\theta) ]}  \log \left|\frac{1-\frac{[1-\sin(\theta)] p}{\tilde{\omega}}}{1+\frac{[1-\sin(\theta) ]p}{\tilde{\omega}}}\right|~.
\end{eqnarray}
Introducing the auxiliary variable $t=1-\sin(\theta)$ we can write
\begin{eqnarray}\label{eq:approximation}
{\rm Re}[\mathcal{J}_1(\omega)]=\int  _{0}^{2}\frac{dt}{\sqrt{t(2-t)}} \frac{1}{\tilde{\omega} t}  \log \left|\frac{\tilde{\omega}-t}{\tilde{\omega}+t}\right|~. 
\end{eqnarray}	
We now note that the integrand in the previous equation is peaked at $t\sim 0 $ and $t\sim \tilde{\omega}$. 
Furthermore, for $\omega \ll 2E_{\rm F}$, both peaks collapse at $t\sim 0$. We can therefore approximate (\ref{eq:approximation}) as
\begin{eqnarray}\label{eq:J1_approx}
{\rm Re}[\mathcal{J}_1(\omega \ll 2E_{\rm F} )] \simeq \int  _{0}^{2}\frac{dt}{\sqrt{2 t}} \frac{1}{\tilde{\omega} t}  \log \left|\frac{\tilde{\omega}-t}{\tilde{\omega}+t}\right|\,,
\end{eqnarray}
which can be evaluated analytically, yielding
\begin{eqnarray}\label{eq:J1_approx2}
&\int  _{0}^{2}\frac{dt}{\sqrt{2 t}} 
\frac{1}{\tilde{\omega} t}  \log \left|\frac{\tilde{\omega}-t}{\tilde{\omega}+t}\right|= & \nonumber \\
&-\frac{\sqrt{\tilde{\omega}} \log \left(\frac{2-\tilde{\omega}}{2+\tilde{\omega}}\right)+2 \sqrt{2} 
\tanh^{-1}(\sqrt{2/\tilde{\omega}})+2 \sqrt{2} \tanh ^{-1}(\sqrt{\tilde{\omega}/2})}{\tilde{\omega}^{3/2}}~.& \nonumber \\
\end{eqnarray}
A further asymptotic expansion in the limit $\omega \ll 2E_{\rm F}$ yields 
\begin{eqnarray}\label{eq:J1exp}
{\rm Re}[\mathcal{J}_1(\omega \ll 2E_{\rm F} )]\simeq -\frac{\sqrt{2} \pi }{\tilde{\omega}^{3/2}}\,.
\end{eqnarray}	
Similarly, we find
\begin{equation}\label{eq:J2exp}
 {\rm Re}[\mathcal{J}_2(\omega \ll 2E_{\rm F} )] \simeq-\frac{\sqrt{2} \pi }{\tilde{\omega}^{3/2}} + \frac{\pi}{\sqrt{2 \tilde{\omega}}}~,
\end{equation}
while, in the same limits, $ {\rm Re}[\mathcal{J}_3(\omega \ll 2E_{\rm F} )]$ is a regular function having a negligible contribution with respect to ${\rm Re}(\mathcal{J}_1) $ and $ {\rm Re}(\mathcal{J}_2)$. Replacing Eqs.~(\ref{eq:J1exp})-(\ref{eq:J2exp}) in Eq.~(\ref{eq:chi0AB_SOM}) we finally obtain
\begin{eqnarray}
{\rm Re}[\chi^{(0)}_{\rm AB}(q_z,q_z^\prime,\vect{q}_\parallel,\omega)]\to -\frac{{ q_x^2+q_z q_z^\prime }}{3L_z\pi^2E_{\rm F}}
\frac{1}{\sqrt{2\tilde{\omega}}}~, 
\end{eqnarray}
which gives Eq.~(14) of the main text. 

In order to evaluate the imaginary part of the response function we exploit the Dirac identity $ {1}/\big[{x^2-(\omega+i0^+)^2}\big]={1}/\big({x^2-\omega^2}\big)+{i\pi}\delta({x^2-\omega^2})$.  
For $\omega<2 E_{\rm F}$ we obtain
\begin{eqnarray}
{\rm Im}[\mathcal{J}_1(\omega)]&=&\int  d p_y \int _0^\infty dp_\perp \Theta(1 - p)\delta\Big(p_y-p+\tilde{\omega} \Big) \frac{\pi}{ \tilde{\omega}p} \,, \nonumber \\ \\
{\rm Im}[\mathcal{J}_2(\omega)]&=&\int  d p_y \int _0^\infty dp_\perp \Theta(1 - p_y)\delta\Big(p_y-p+\tilde{\omega} \Big) \frac{\pi}{ \tilde{\omega}p}\,, \nonumber \\ \\
{\rm Im}[\mathcal{J}_3(\omega)]&=&0\,,
\end{eqnarray}
which, replaced in Eq.~(\ref{eq:chi0AB_SOM}), result into
\begin{eqnarray}
{\rm Im}[\chi^{(0)}_{\rm AB}(q_z,q_z^\prime,\vect{q}_\parallel,\omega)]=-\frac{{ q_x^2+q_z q_z^\prime }}{3L_z\pi^2E_{\rm F}}
\frac{\big(\sqrt{2+\tilde{\omega}}-\sqrt{2-\tilde{\omega}}\big)}{\tilde{\omega}^\frac{3}{2}} \,.\nonumber \\ 
\end{eqnarray}
The leading term for $\omega \ll E_{\rm F}$ gives Eq.~(23) of the main text. 

Following similar steps in the undoped case ($E_{\rm F}=0$), we find
\begin{equation}
\chi^{(0)}_{\rm AB}(q_z,q_z^\prime,\vect{q}_\parallel,\omega)\to-i \frac{2\big( q_x^2+q_z q_z^\prime \big)}{3L_z\pi^2 \omega} \,.
\end{equation}
%

%


%

\end{document}